\documentclass{pasj00}
\usepackage{lscape}

\begin{document}
\SetRunningHead{T.~Inui et al.}{Chandra Observation of the Starburst Galaxy NGC~2146}
\Received{2004/05/06}%{yyyy/mm/dd}
\Accepted{2004/10/29}%{yyyy/mm/dd}

\title{Chandra Observation of the Starburst Galaxy NGC~2146}

%%% begin:list of authors
\author{
	Tatsuya \textsc{Inui}\altaffilmark{1},
	Hironori \textsc{Matsumoto}\altaffilmark{1},
	Takeshi Go \textsc{Tsuru}\altaffilmark{1},
	Katsuji \textsc{Koyama}\altaffilmark{1},
	\\
	Satoki \textsc{Matsushita}\altaffilmark{2},
	Alison B. \textsc{Peck}\altaffilmark{3},
	\and
	Andrea \textsc{Tarchi}\altaffilmark{4} \altaffilmark{5}
}
\altaffiltext{1}{Department of Physics, Graduate School of Science, Kyoto University, Sakyo-ku, Kyoto 606-8502, Japan;}
\email{inuit@cr.scphys.kyoto-u.ac.jp}
\email{matsumoto@cr.scphys.kyoto-u.ac.jp}
\email{tsuru@cr.scphys.kyoto-u.ac.jp}
\email{koyama@cr.scphys.kyoto-u.ac.jp}
\altaffiltext{2}{Academia Sinica, Institute of Astronomy and Astrophysics, P.O.Box 23-141, Taipei 106, Taiwan, R.O.C.}
\email{satoki@asiaa.sinica.edu.tw}
\altaffiltext{3}{Harvard-Smithsonian Center for Astrophysics, 60 Garden St., MS-78, Cambridge, MA 02138, U.S.A.}
\email{apeck@cfa.harvard.edu}
\altaffiltext{4}{Instituto di Radioastronomia, CNR, Via Gobetti 101, 40129 Bologna, Italy}
\altaffiltext{5}{Osservatorio Astronomico di Cagliari, Loc. Poggio dei Pini, Strada 54, 09012 Capoterra (CA), Italy}
\email{atarchi@ira.cnr.it}
%%% end:list of authors

%%% Please use the following style in case that sorting by 
%%% affilation is impossible. 
%
% \author{
%   D-Firstname \textsc{D-Familyname}\altaffilmark{1}
%   E-Firstname \textsc{E-Familyname}\altaffilmark{1,2}
%   and
%   F-Firstname \textsc{F-Familyname}\altaffilmark{2}}
% \altaffiltext{1}{Address of Institute}
% \email{ddddd@xxx.xxx.xx.xx}
% \email{eeeee@xxx.xxx.xx.xx}
% \altaffiltext{2}{Address of Institute}

%% `\KeyWords{}' always has to be placed before `\maketitle'.
 %Do NOT move this preamble from here!

\maketitle

\begin{abstract}

We present six monitoring observations of the starburst
galaxy NGC~2146 using the Chandra X-ray Observatory. We have
detected 67 point sources in the $\timeform{8'.7} \times
\timeform{8'.7}$ field of view of the ACIS-S detector.
Six of these sources were Ultra-Luminous X-ray Sources, 
the brightest of which has a luminosity of $5 \times 10^{39}$
 ergs s$^{-1}$. One of the source, with a luminosity of $\sim 1 \times 10^{39}$ ergs s$^{-1}$, is coincident with the dynamical center location, as derived from the $^{12}$CO rotation curve. We suggest that this source may be a low-luminosity active galactic nucleus. We have produced a table where the positions and main characteristics of the Chandra-detected sources are reported. The comparison between the positions of the X-ray sources and those of compact sources detected in NIR or radio does not indicate any definite counterpart. Taking profit of the relatively large number of sources detected, we have derived a $\log{N}-\log{S}$ relation and a luminosity function. The former shows a break at $\sim 10^{-15}$ ergs cm$^{-2}$ s$^{-1}$, that we interpret as due to a detection limit. The latter has a slope above the break of 0.71, which is similar to those found in the other starburst galaxies. In addition, a diffuse X-ray emission has been detected in both, soft (0.5--2.0~keV) and hard (2.0--10.0~keV), energy bands. The spectra of the diffuse component has been fitted with a two (hard and soft) components. The hard power-law component, with a luminosity of $\sim 4\times 10^{39}$~ergs~s$^{-1}$, is likely originated by unresolved point sources, while the soft component is better described by a thermal plasma model with a temperature of 0.5~keV and high abundances for Mg and Si.

\end{abstract}

\section{Introduction}

ASCA observations have shown that X-ray spectra of starburst galaxies generally possess a hard component above 2 keV (\cite{Dahlem1998}, \cite{Ptak1999}). However, the origin of such a component is still unclear. In the prototype starburst galaxy M82, we found that most of the hard component is due to the most luminous X-ray source, M82~X-1, and that this object may belong to a new black holes class of objects, with $10^3-10^6 \MO$, named Intermediate-Mass Black
Holes (IMBHs; \cite{Matsumoto2001}). In another well-studied starburst galaxy,
NGC~253, it is hot plasma with a temperature of 6~keV that accounts for the hard component \citep{Pietsch2001}.

Recently, we have studied 13 starburst galaxies previously observed with ASCA
(\cite{Dahlem1998}, \cite{Ptak1999}). Among these we have found that
NGC~2146 is the object with the high ratio of 2-10~keV X-ray to B-band luminosity ($\log{L_{\rm X}/L_{\rm B}}=30.38$~ergs~s$^{-1}$~$L_{\rm B,\MO}^{-1}$). In order to account for such a high ratio it is necessary for the presence of a large number (and maybe a different kind) of X-ray sources of emission than the usual X-ray binaries. This fact renders NGC~2146 the most interesting target in the sample to investigate the origin of the hard X-ray component.

NGC~2146 is a nearby ($D$=11.6~Mpc, $\timeform{1'}\sim 3.5$~kpc; \cite{DellaCeca1999}) SB(s)ab, edge-on ($i \sim 63$\r{}) starburst galaxy. The optical size of the galaxy ($D_{\rm 25}$ ellipse; at the limiting surface brightness of 25 $B$ mag arcsec$^{-2}$ after correction for Galactic extinction) is about $\timeform{6'.0}\times\timeform{3'.4}$ \citep{Vaucouleurs1991}. The dynamical center of the galaxy lies toward a dense dust lane \citep{Young1988}. NGC~2146 has an outflow of hot gas along the minor axis driven by supernova explosions and stellar winds in the starburst region (\cite{Armus1995}, \cite{DellaCeca1999}, \cite{Greve2000}). X-ray luminosities of NGC~2146 derived from the ASCA observation were $\sim 1.3 \times 10^{40}, \sim 1.8 \times 10^{40},$ and
$\sim 3.1 \times 10^{40}$~ergs~s$^{-1}$ in the soft
(0.5--2.0~keV), hard (2.0--10.0~keV), and total
(0.5--10.0~keV) energy bands,
respectively\citep{DellaCeca1999}.

The unprecedented spatial resolution (\timeform{0".5}) of Chandra makes it the most suitable instrument to determine whether the hard X-ray component of
NGC~2146 is due to point sources such as IMBHs or to a hot
diffuse plasma.

In this paper, we present the detailed catalog of point
sources detected from the monitoring observations of
NGC~2146 with Chandra and discuss their physical natures. Furthermore, we investigate the diffuse emission component and try to shed a light on the origin of the hard X-ray component. Errors and uncertainties in this paper refer
to 90\% confidence limits ($\Delta\chi^2$=2.706) unless otherwise stated.

\section{Observations}

NGC~2146 was observed six times through August to December
2002, with the Advanced CCD Imaging Spectrometer 
(ACIS; \cite{Garmire2003}) on board the Chandra X-ray
Observatory (CXO)\citep{Weisskopf2002}. The exposure time of each observation was about 10~ks, yielding a total observing time of 60~ks. Observations dates, exposure times and background levels are listed in Table~\ref{TBL:OBS_LOG}. The nominal position of ACIS-S3 (a back-illuminated CCD on the spectroscopic array (ACIS-S) with good charge-transfer efficiency and good quantum efficiency below 0.5~keV) was selected coincident with the body of NGC~2146.

\begin{table*}
\begin{center}
\caption{Log of the observations}\label{TBL:OBS_LOG}
\begin{tabular}{lccc} 
\hline\hline
Observation ID & Date & Exposure & Background\\
\multicolumn{1}{c}{} & (yyyy/mm/dd) & (s) & (counts s$^{-1}$ arcmin$^{-2}$) \\
\hline
3131.............. & 2002/08/30 & 9,276 & 0.012 \\
3132.............. & 2002/09/20 & 9,523 & 0.013 \\
3133.............. & 2002/10/09 & 9,832 & 0.081 \\
3134.............. & 2002/10/27 & 9,576 & 0.012 \\
3135.............. & 2002/11/16 & 10,020 & 0.013 \\
3136.............. & 2002/12/05 & 9,835 & 0.014 \\
Total............. &  & 58,062 & \\
\hline
\end{tabular}
\end{center}
\end{table*}

\section{Analysis and Results}
\subsection{Imaging}
\begin{figure*}
(a)\hspace{0.5\textwidth}(b)\\
\vspace{2em}
  \begin{center}
    \FigureFile(80mm,80mm){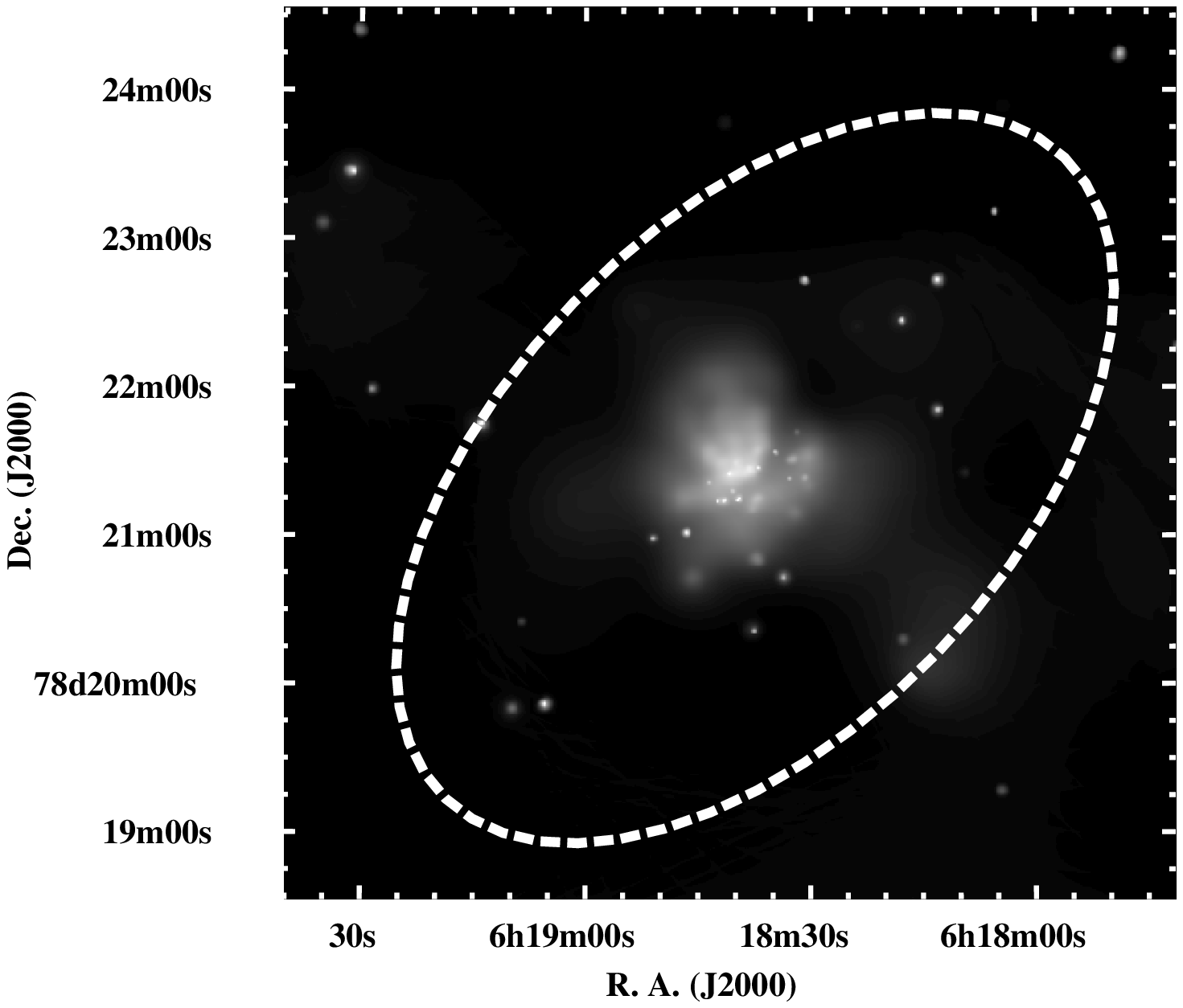}
    \FigureFile(80mm,80mm){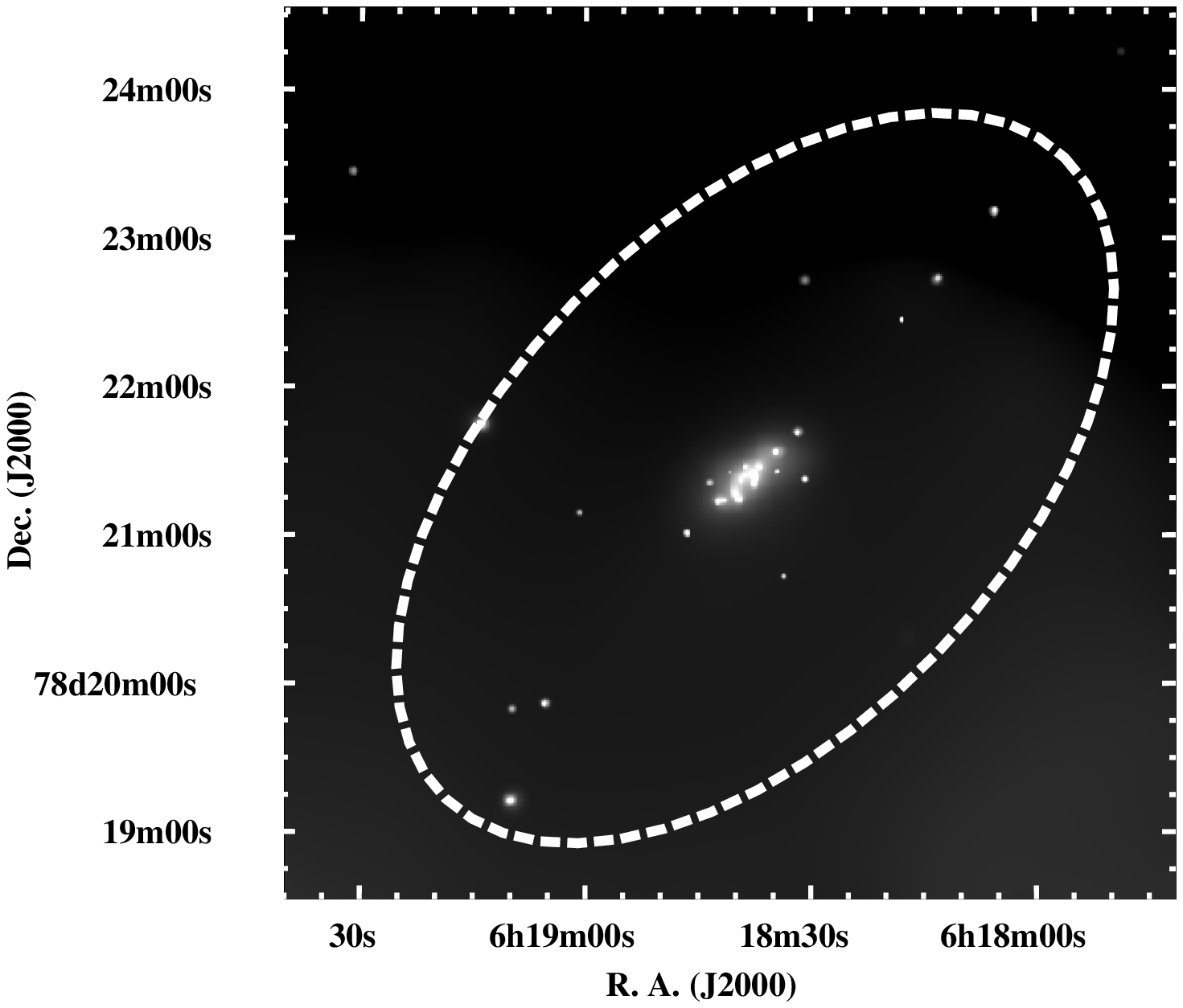}    
    %%% \FigureFile(width,height){filename}
  \end{center}
  \caption{ Chandra ACIS-S $\timeform{6'.0} \times \timeform{6'.0}$
  images of NGC~2146 in (a) the 0.3--2.0~keV and (b) the 2.0--10.0~keV
  energy bands. Each image is the combined image of the six observations
  and is adaptively smoothed. White dashed ellipses show
  $D_{\rm 25}$ ellipses ($\timeform{6'.0}
  \times \timeform{3'.4}$) of NGC~2146.}
  \label{FIG:ACIS-S3}
\end{figure*}

For the following discussion it is worth to emphasize that, as explained in detail in the next section, there is no appreciable positional offsets among the observations performed at six different epochs. Hence, we were legitimate to combine the images. Using $dmimgcalc$ and $csmooth$ program in the CIAO package version 2.3\footnote{See http://asc.harvard.edu/ciao/}, we produced two adaptive smoothed maps of NGC~2146  in the two energy bands: 0.3--2.0~keV (Figures~\ref{FIG:ACIS-S3}(a)) and 2.0--10.0~keV (Figures~\ref{FIG:ACIS-S3}(b)), respectively. When compared to the optical
image of the galaxy, the X-ray emission is found to be concentrated toward the
nuclear region of NGC~2146. Shown in Figure~\ref{FIG:ACIS-S3}(a) and Figure~\ref{FIG:ACIS-S3}(b), the soft diffuse emission brows off from the galactic plane, while the hard diffuse radiation lies along the galactic plane.

\subsection{Point Sources}

\subsubsection{Source Detection}

In order to establish the extent of the positional offsets between our multi-epoch Chandra maps, we run the $wavdetect$ program in the CIAO package on
the 0.3--10.0~keV image of each epoch.

Wavelet scales were from 1 to 16 pixels in multiples of $\sqrt{\mathstrut 2}$. We found that among the six observations the positions of the two brightest sources are consistent within \timeform{0".2}, which is comparable
to the Chandra positional accuracy of \timeform{0".1}
\footnote{See http://cxc.harvard.edu/cal/}.

Images at different epochs can be combined to increase signal-to-noise ratio
and allow the detection of faint sources. Then, we run the $wavdetect$ program
on the combined images for three energy bands: total
(0.3--10.0~keV), soft (0.3--2.0~keV) and hard (2.0--10.0~keV)
energy bands, obtaining a detected source number of 62, 55 and 42, respectively. Since five sources are detected only in the soft (three sources) or hard (two sources) energy bands, we have a total number of 67~X-ray sources in the $\timeform{8'.7} \times \timeform{8'.7}$ ACIS-S field of view (FOV).
41 point sources are within the $D_{\rm 25}$ ellipse at a $5\sigma$ level.
We considered that these 41 sources belong to NGC 2146 and produce a $\log(N)-\log(S)$ distribution and a luminosity function for NGC~2146. Among these sources, about five sources are expected to be background sources which are within the $D_{\rm 25}$ ellipse by chance \citep{Giacconi2001}.

In our analysis we have adopted the following assumption: the source regions' extent has been taken to be two times the standard deviation of the point spread function (PSF) at the detected positions. In most cases, as the background region, we used an ellipse twice as large as that of the source, excluding the source region. Instead at the galactic center the background region was selected close to the source because it was impossible to do otherwise due to the crowding of the field.

\subsubsection{Timing Analysis}
\begin{figure*}
  \begin{center}
    \FigureFile(50mm,40mm){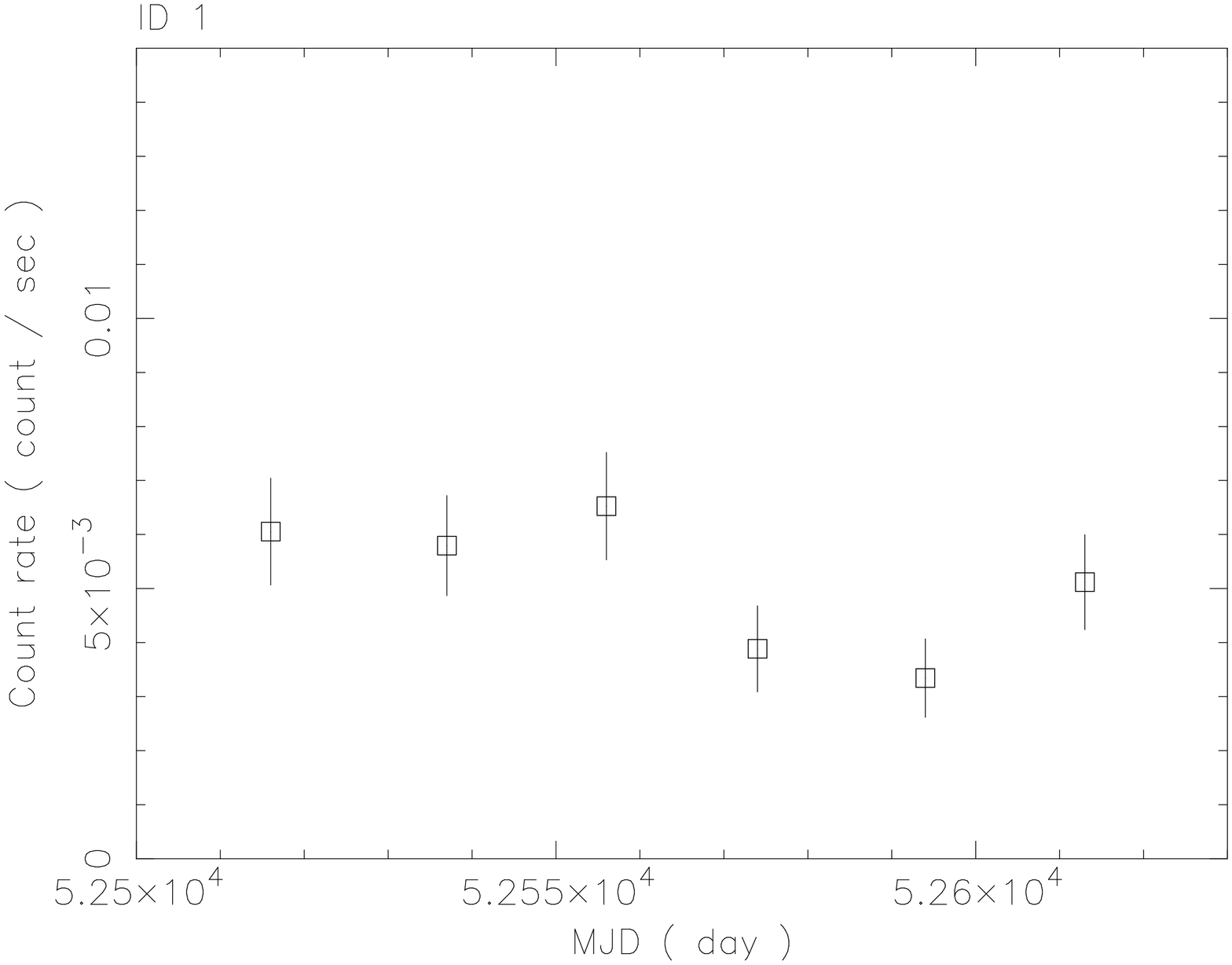}
    \FigureFile(50mm,40mm){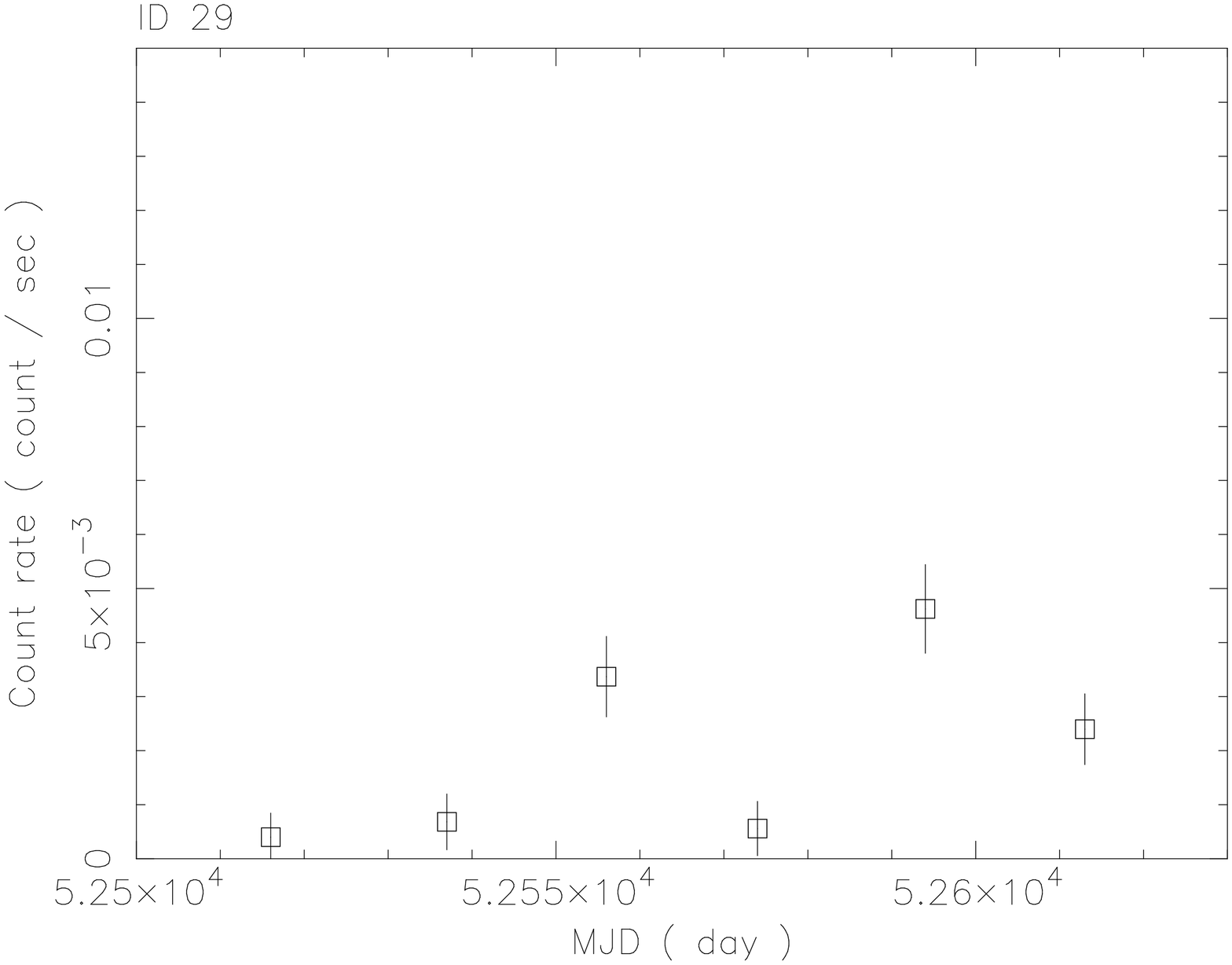}
    \FigureFile(50mm,40mm){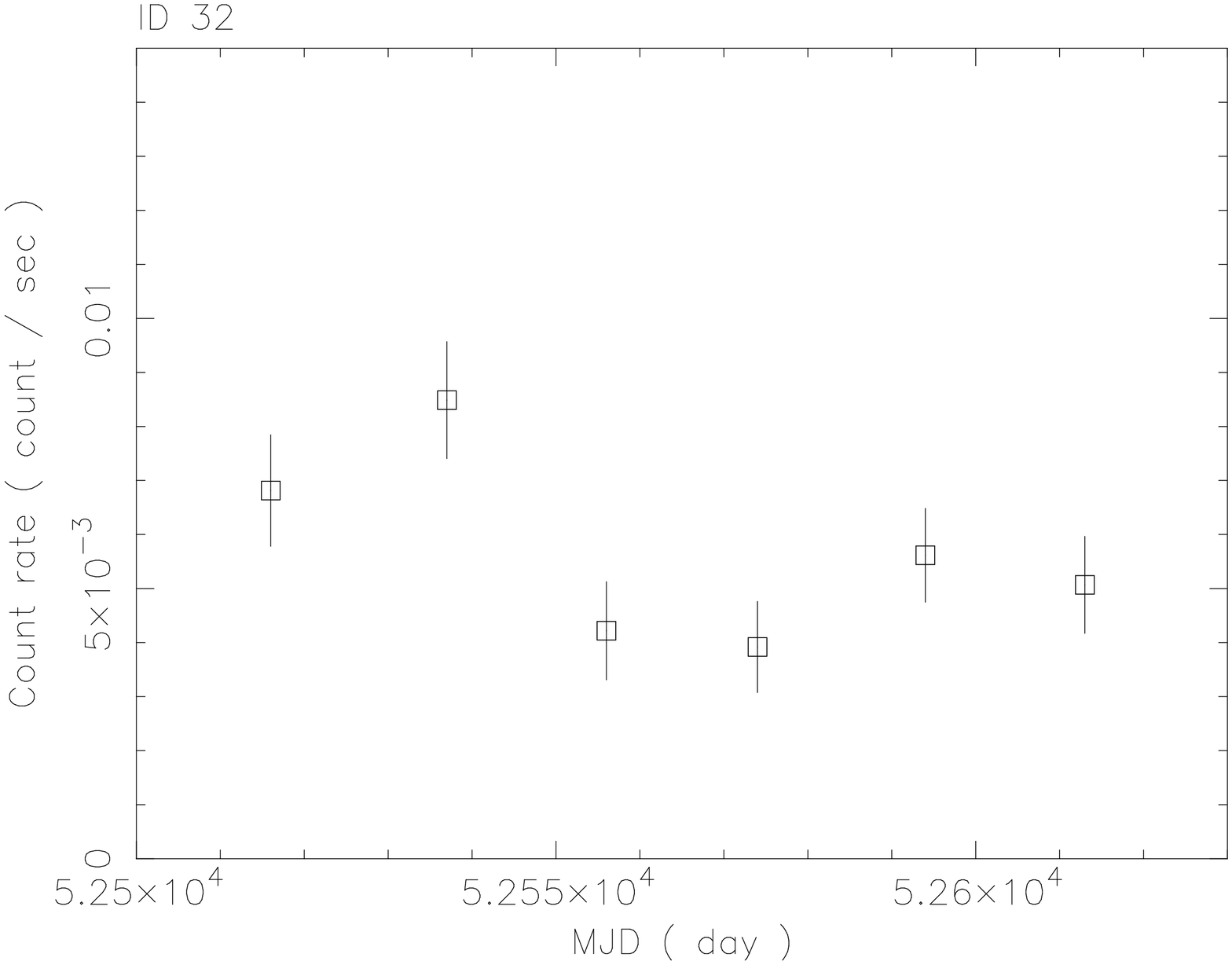}
    \FigureFile(50mm,40mm){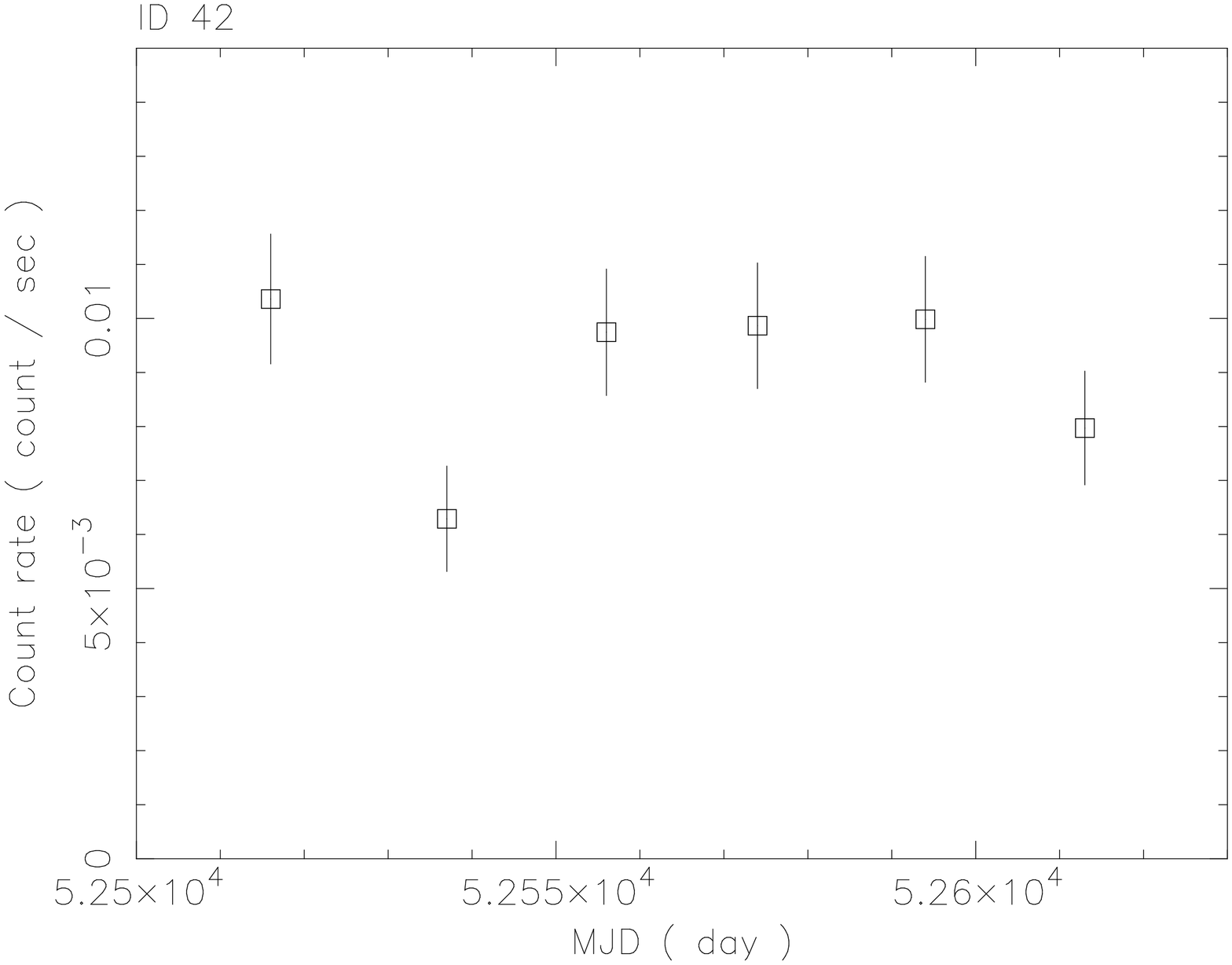}
    \FigureFile(50mm,40mm){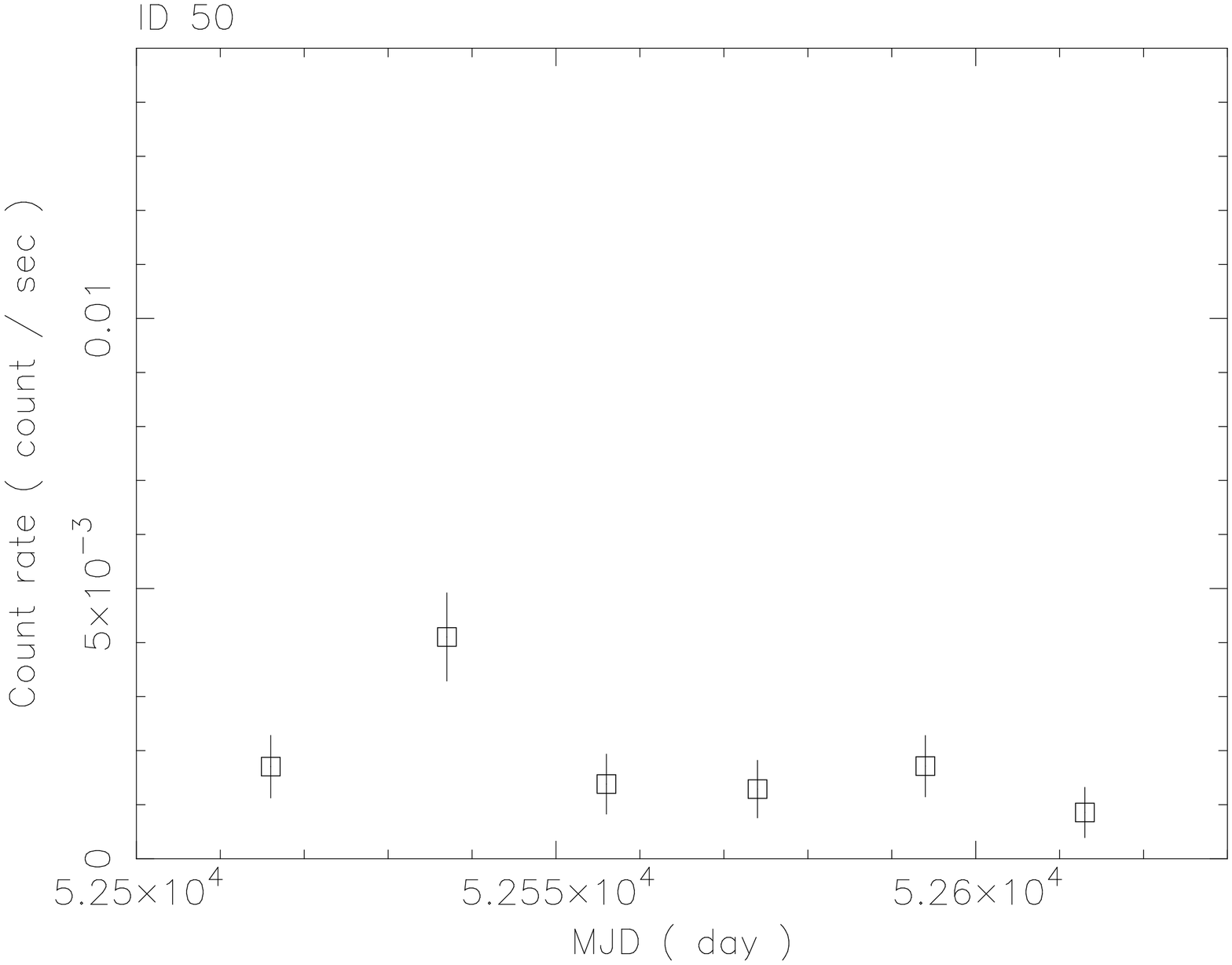}
    %%% \FigureFile(width,height){filename}
  \end{center}
  \caption{ Light curves of five time-variable sources in
the 0.5-10.0 keV band. Error bars indicate
$1\sigma$.}\label{FIG:LC15+16}
\end{figure*}

Light curves of the point sources have been created using the mean
count rate of each observation and fitted with a constant count rate model.
For five sources we obtained reduced $\chi^2$ values larger than 11.05/5(dof), indicating, with a confidence level more than 95\%, a significant variability with time.
The light curves for these five variable
sources are shown in Figure~\ref{FIG:LC15+16}.

\subsubsection{Spectral Analysis}
We extracted the spectrum of each point source at each
epoch and analyzed it with XSPEC version 11.0.1 in
XANADU\footnote{See http://heasarc.gsfc.nasa.gov/docs/software/lheasoft/}.
In order to use $\chi^2$ statistics,  we grouped the energy bins of the
spectra so that each bin contains at least twenty counts.
We fitted the spectra simultaneously with
an absorbed power-law model.
For each epoch, we used common values of the absorption column density $N_{\rm H}$ and the photon index $\Gamma$, while the normalization was left as a free parameter.

For bright sources having more than 180 total counts in the six
observations, all parameters, $N_{\rm H}$, $\Gamma$ and
normalizations, were free. For sources having less than 180
total counts, we fixed $\Gamma$ to 2.0. For the faintest
sources whose counts are less than 100 total counts, we
combined the six spectra and fitted it with $\Gamma=2.0$ and
$N_{\rm H} = 4.0 \times 10^{21}$ cm$^{-2}$, which is the
best-fit $N_{\rm H}$ of the diffuse emission (see below).
The observed flux and absorption corrected luminosity of the point sources
in the 2.0--10.0~keV band and the 0.5--10.0~keV band are shown
in the source list (Table~\ref{TBL:CATALOG}). 
Seven sources (ID29,32,33,35,37,42,60) have luminosities of
over $10^{39}$~ergs~s$^{-1}$, and all except ID60 are within the $D_{\rm 25}$
ellipse and considered to be Ultra-Luminous X-ray Sources (ULXs). ID60 is so
remote from NGC~2146 that it should be a background AGN.

\subsubsection{Diffuse Source}
We extracted the spectra of the diffuse emission from a
circular region at the center of NGC~2146 with a radius of
\timeform{1'.8} excluding the inner point sources. The
background spectrum for each observation was extracted from
a source-free rectangular region (about 5 arcmin$^2$ area)
around the position (detX,detY)=(830,830) in detector coordinates. The
background level of the third observation was
0.081~counts~s$^{-1}$~arcmin$^{-2}$ and quite variable with
an amplitude of 0.03~counts~s$^{-1}$~arcmin$^{-2}$, while
those of the other observations were $\sim
0.01$~counts~s$^{-1}$~arcmin$^{-2}$ and stable during the
observations.
This effect can be ignored in the case of point sources while not in the case of the diffuse component due to the area where counts are accumulated. We therefore did not use the third observation
for the analysis of the diffuse emission.

\begin{figure}
  \begin{center}
    \FigureFile(80mm,90mm){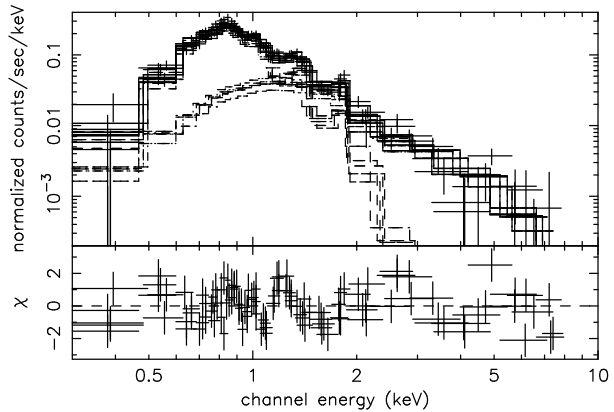}
    %%% \FigureFile(width,height){filename}
  \end{center}
  \caption{
Spectra of the diffuse emission (except the third
observation) and the best-fit thin-thermal plasma plus
power-law model. Features are seen between 1 and 2~keV,
which are considered to come from Mg~XI (1.33--1.35~keV) and
Si~XIII (1.84--1.87~keV) emission lines}\label{FIG:SPEC-DIF}
\end{figure}

The derived spectra for the diffuse component are shown in Figure~\ref{FIG:SPEC-DIF}. We identified the strong spectral features as produced by Mg~XI (1.33--1.35~keV) and Si~XIII (1.84--1.87~keV). This suggests that the emission is originated by optically-thin thermal plasma.
It was not possible to fit the spectra with an absorbed single-temperature plasma model with variable abundances ("vmekal" model; \cite{Mewe1985a}, \cite{Mewe1985b}, \cite{Liedahl1995}). 

Hence, we used instead a combination of a "vmekal" plus a power-law model with the same absorption column density for both models. The equivalent width of Fe K line (6.4~keV) was estimated to be $\le 3$~keV. An additional attempt has been made to fit the spectra with the "vmekal" plus a thermal bremsstrahlung model. The derived parameters for both fits are summarized in Table~\ref{TBL:VMEKAL}.

Still some line-like residuals are found in Figure~\ref{FIG:SPEC-DIF}, however,
we cannot improve $\chi^2$ value to add another thermal component. The origin of this line component will be discussed in \S{4.3.2}.

\begin{table*}
\begin{center}
\caption{Results of the spectral fits of the diffuse emission}\label{TBL:VMEKAL}
\begin{tabular}{lcc}
\hline\hline
Model & absorbed vmekal+power-law & absorbed vmekal+bremsstrahlung \\
\hline
$N_{\rm H}$ ($10^{22} {\rm cm}^{-2}$) & 0.42 (0.31 -- 0.58) & 0.40 (0.27 -- 0.66) \\
\hline
\multicolumn{3}{c}{soft component} \\
\hline
$kT_{\rm soft}$ (keV) & 0.30 (0.24 -- 0.37) & 0.31 (0.26 -- 0.39) \\
Mg\footnotemark[$*$] & 2.3 (1.8 -- 3.0) & 2.2 (1.7 -- 3.1) \\
Si\footnotemark[$*$] & 4.3 (2.0 -- 7.2) & 3.3 (1.2 -- 7.5) \\
Fe\footnotemark[$*$] & 0.80 (0.59 -- 1.06) & 0.73 (0.55 -- 1.05) \\
$F_{\rm 0.5-2.0keV}$\footnotemark[$\dagger$] ($10^{-13}$ ergs cm$^{-2}$ s$^{-1}$) & 4.0 & 4.2 \\
$L_{\rm 0.5-2.0keV}$\footnotemark[$\dagger$] ($10^{40}$ ergs s$^{-1}$) & 3.6 & 3.2 \\
\hline
\multicolumn{3}{c}{hard component} \\
\hline
$\Gamma/kT_{\rm hard}$ (keV) & 2.4 (2.1 -- 2.7) & 2.4 (1.8 -- 3.5) \\
$F_{\rm 2.0-10.0keV}$\footnotemark[$\ddagger$] ($10^{-13}$ ergs cm$^{-2}$ s$^{-1}$) & 2.7 & 2.2 \\
$L_{\rm 2.0-10.0keV}$\footnotemark[$\ddagger$] ($10^{40}$ ergs s$^{-1}$) & 0.45 & 0.37 \\
\hline      
$\chi^2$/dof & 116/91 & 118/91 \\
\hline
\multicolumn{3}{@{}l@{}}{\hbox to 0pt{\parbox{185mm}{\footnotesize
       Note---Parentheses indicate the 90\% confidence limit.
	   \par\noindent
       \footnotemark[$*$] Abundances are relative to the \cite{Anders1989} solar abundance.
	\par\noindent
       \par\noindent
       \footnotemark[$\dagger$] Observed flux and absorption-corrected luminosity of the soft component in the 0.5-2.0 keV band.
	\par\noindent
       \footnotemark[$\ddagger$] Observed flux and absorption-corrected luminosity of the hard component in the 2.0-10.0 keV band.
     }\hss}}
\end{tabular}
\end{center}
\end{table*}

\section{Discussion}

\subsection{Comparison with the ASCA results}

\begin{figure}
  \begin{center}
    \FigureFile(80mm,90mm){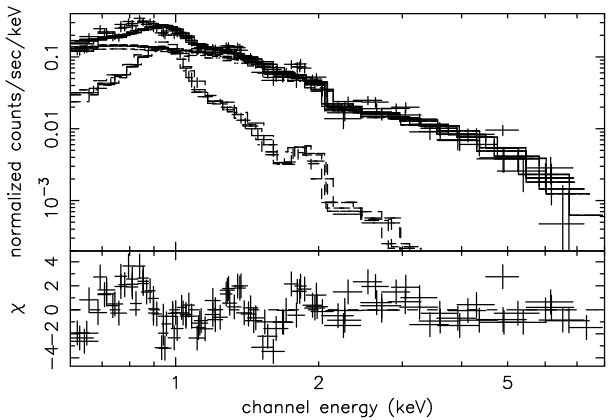}
    %%% \FigureFile(width,height){filename}
  \end{center}
  \caption{ Spectra of the entire NGC~2146 (except the third
observation) and the ASCA two-component (thin-thermal plasma
plus power-law) model. Large excess are found below
1~keV.}\label{FIG:SPEC-OVAL-FIX}
\end{figure}

In order to compare our Chandra data with the previous ASCA results
\citep{DellaCeca1999}, we extract
the entire NGC~2146 spectra from a circle at the center of the galaxy
with a radius of \timeform{1'.8} including the inner point
sources. Since we spatially resolved point sources, we operate a
simple spectral analysis to examine the overall flux variability.

We ignore those spectra with energies below 0.6~keV because in this range ASCA has no effective area. We use the same fitting model used by \citet{DellaCeca1999}, an optically-thin thermal plasma model plus a power-law model.
At first, all parameters except normalizations are fixed at
the ASCA best-fit values, $N_{\rm H}=7 \times 10^{20}~{\rm cm^{-2}}$,
the plasma temperature of $kT=0.82$~keV, solar abundance and $\Gamma=1.7$. 
Figure~\ref{FIG:SPEC-OVAL-FIX} shows the
Chandra spectra of NGC~2146 together with the ASCA model and residuals
after the model is applied. It is noticeable that the model causes large
residuals below 1~keV. We believe that this excess is real, since, below 1~keV, the Chandra ACIS-S has a better effective area than ASCA.
Next, we fit them with all parameters free. The result is reported
in Table~\ref{TBL:DIF-PARAM}. No significant difference in flux
between two models is found while the residuals in the low energy band are
improved. The power-law index $\Gamma$ and the observed flux
is comparable with the ASCA results.
Therefore we conclude that the X-ray emission from the entire region of
NGC~2146 shows no significant flux variability between our Chandra
and the ASCA observations.

\begin{table*}
\begin{center}
\caption{Comparison with the ASCA result}\label{TBL:DIF-PARAM}
\begin{tabular}{ccccccc}
\hline\hline 
$N_{\rm H}$ & $kT_{\rm soft}$ & Abundance & $\Gamma$ & $F_{\rm 0.5-2.0}$ & $F_{\rm 2.0-10.0}$ & $\chi^2$/dof \\
(1) & (2) & (3) & (4) & (5) & (6) & (7) \\\hline
0.07 (fixed) & 0.82 (fixed) & 1.0 (fixed) & 1.7 (fixed) & 5.9 & 9.9 & 260/110 \\0.15 (0.10--0.20) & 0.59 (0.55--0.63) & 0.63 (0.10--5.24) & 1.7 (1.5--1.9) & 5.8 & 9.8 & 174/106 \\
\hline
\multicolumn{7}{@{}l@{}}{\hbox to 0pt{\parbox{185mm}{\footnotesize
       Note---Parentheses indicate the 90\% confidence limit.
       Col. (1): Absorption column density ($10^{22} \rm cm^{-2}$).
       Col. (2): Plasma temperature (keV).
       Col. (3): Abundance (solar abundance).
       Col. (4): Photon index.
       Col. (5) and (6): Observed flux of the soft component in the 0.5-2.0 keV band and the 2.0-10.0 keV band ($10^{-13}$ ergs cm$^{-2}$ s$^{-1}$).
       Col. (7): $\chi^2$ and degree of freedom.
     }\hss}}
\end{tabular}
\end{center}
\end{table*}

\subsection{Point Sources}

\subsubsection{Search for an IMBH}
The IMBH candidate M82~X-1 is a
very hard source with a luminosity of over $10^{41}$ ergs
s$^{-1}$, flux variation on a time scale of $\sim
10^4$~s and off-center position
(\cite{Matsumoto1999},\yearcite{Matsumoto2001}). In our
observations of NGC~2146, six ULXs are found
and three of them (ID29,32,42) show time variations.
The luminosity of the highest
count-rate source ID42 is $2 \times 10^{39}$~ergs~s$^{-1}$ but
it is a rather soft (HR $\sim -0.14$) source. On
the other hand, ID29 and ID32 are hard sources
(HR $\sim$ 0.55 and 0.44, respectively) with
luminosities of $3 \times 10^{39}$ ergs s$^{-1}$ and $4
\times 10^{39}$ ergs s$^{-1}$, respectively. The light
curve of ID29 (Figure~\ref{FIG:LC15+16}) has two flare-like
peaks and the count rate at the maximum peak is about three times
larger than the average count rate. We find no point source
whose luminosity is comparable to that of M82 X-1.

\subsubsection{Identifications of the Chandra sources}
In order to understand the nature of each point source, we
searched for the NIR and radio counterparts using
the 2MASS All-Sky Point Source
Catalog(PSC)\footnote{See http://www.ipac.caltech.edu/2mass/} and
the results of MERLIN+VLA observations~\citep{Tarchi2000} and
found no positional coincidence between the
detected X-ray point sources and those found in the NIR or radio
band. The correlation method is explained in the following paragraph.

\begin{figure*}
  \begin{center}
    \FigureFile(60mm,70mm){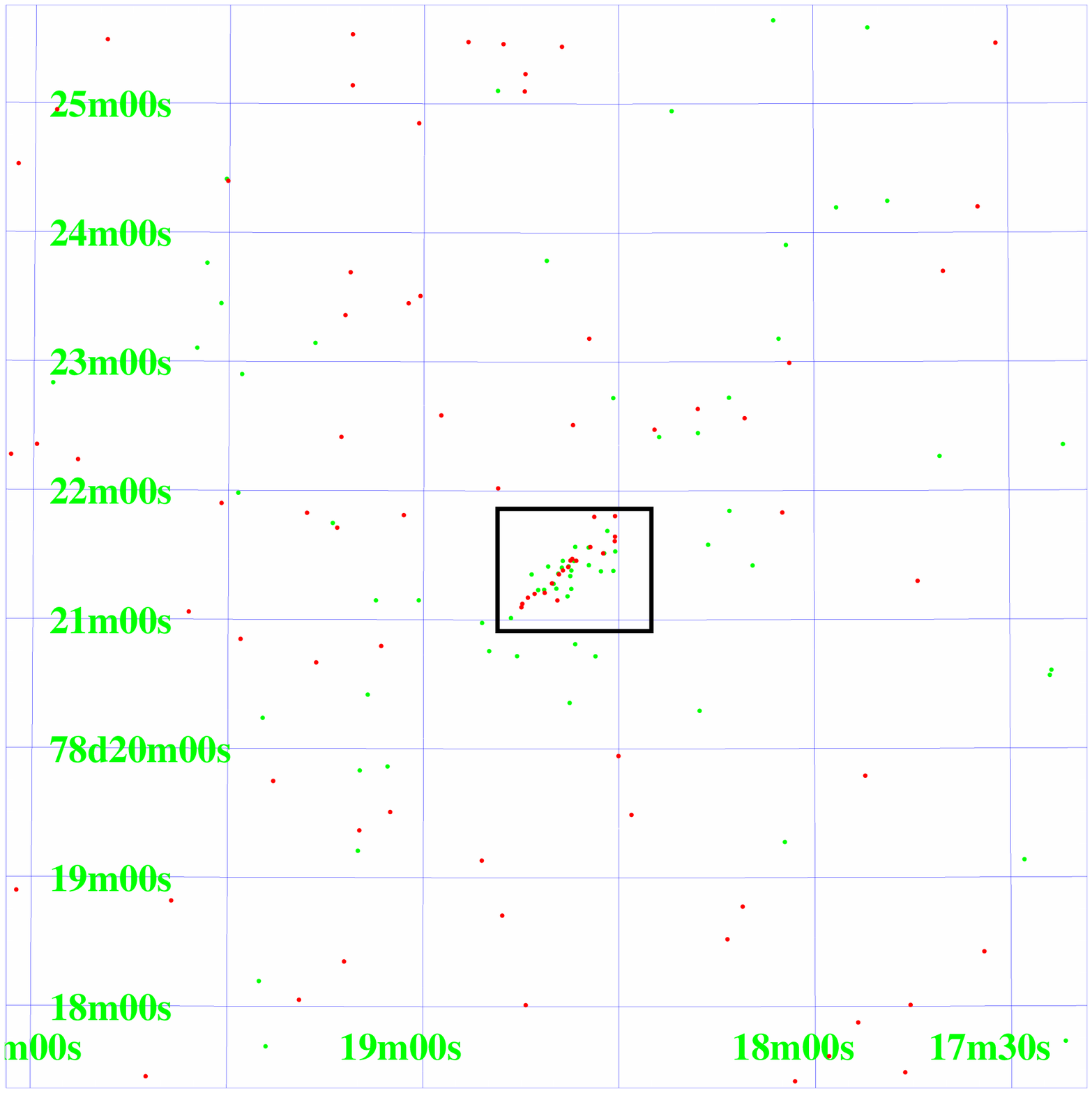}
    \FigureFile(60mm,60mm){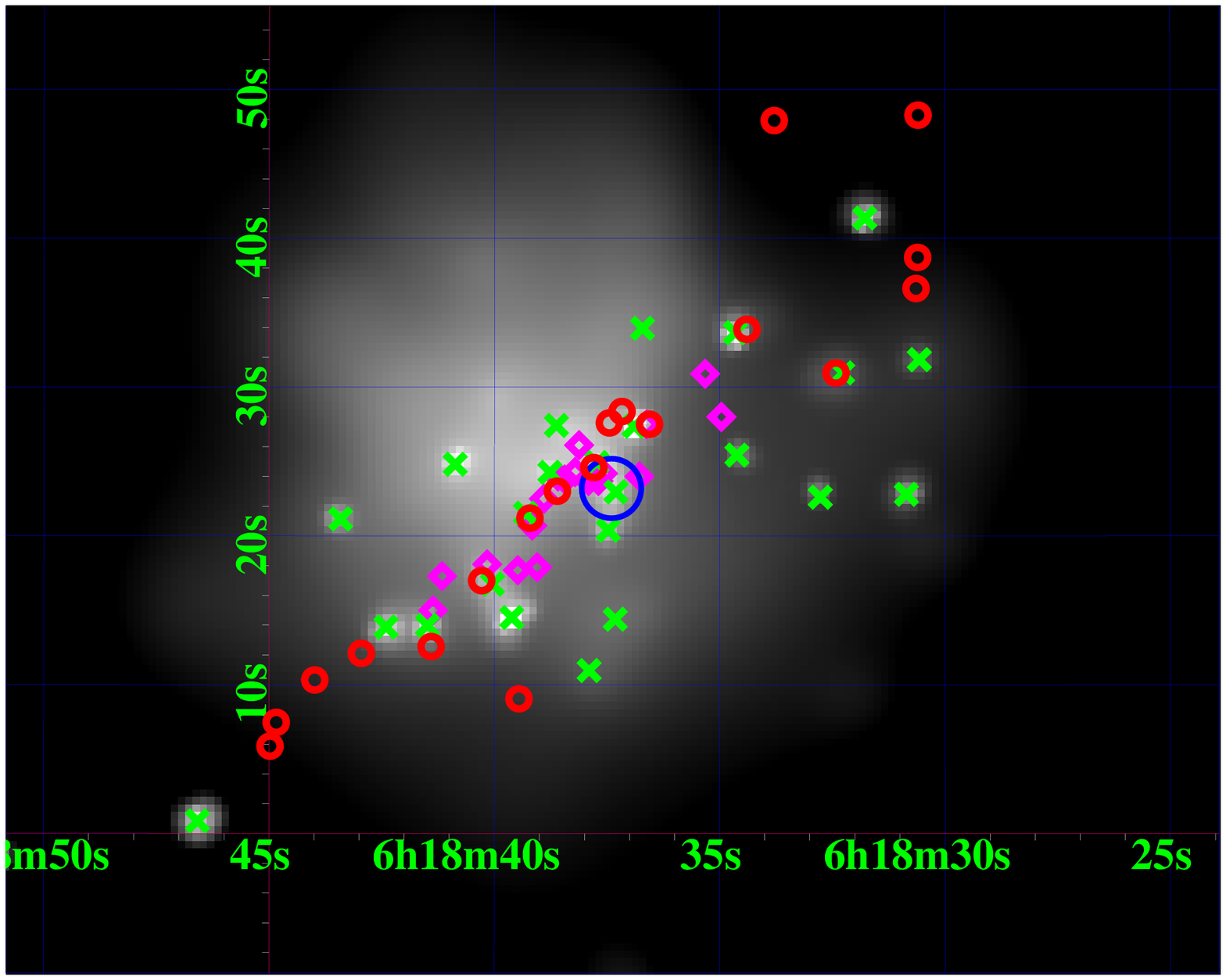}
    %%% \FigureFile(width,height){filename}
  \caption{ Spatial distributions of the Chandra (green mark,
cross), the 2MASS (red mark, circle) and the MERLIN and the VLA
(magenta mark, diamond) sources. The magnified image around
the galactic center (rectangle area in the left image) is
shown in the right. A blue circle indicates the dynamical
center.}\label{FIG:DIST-2MASS}
  \end{center}
\end{figure*}

Figure~\ref{FIG:DIST-2MASS} shows the
distribution of NIR sources overlaid on the Chandra
image. We picked up the closest NIR source from each X-ray
source and did vice versa. Forty pairs were commonly
selected in both the methods. Among the pairs, we found that
three pairs have a distance of less than \timeform{0".5}
between the X-ray and the NIR sources, and we found no pairs
whose distance was less than \timeform{0".2}. Next we
checked the absolute positional offset of the ACIS-S
coordinate. Using comparably nearby 10 pairs within
\timeform{3".0}, we shifted the ACIS-S frame to minimize the
total distances of these pairs and searched pairs
again. However, there were the same three pairs within
\timeform{0".5}, and were no pairs within \timeform{0".2}. We
then decided to use the original ACIS-S coordinate. The
positional accuracy of Chandra is \timeform{0".1}
\footnote{See http://cxc.harvard.edu/cal/Hrma/optaxis/platescale/geom\_public.html for ACIS-S positional accuracy},
while that of 2MASS sources is $\sim 80$ mas
\footnote{See http://www.ipac.caltech.edu/2mass/releases/allsky/doc/sec1\_6b.html for the 2MASS PSC Position Reconstruction}.
Therefore we concluded that there were no
identifications with the NIR sources. We also searched
for the radio counterparts using the results of the MERLIN and
the VLA observations~\citep{Tarchi2000} in the same way as
for the NIR identification. As the positional uncertainties
of the MERLIN and the VLA are less than \timeform{0".1}
\footnote{See http://www.merlin.ac.uk/user\_guide/OnlineMUG/newch0-node104.html for the MERLIN Position accuracy, and http://www.vla.nrao.edu/astro/guides/vlas/current/node20.html for the VLA Positional Accuracy},
we used the same criterion as 2MASS. We found no radio
counterparts, either.

\subsubsection{The galactic center source}

We find the X-ray source ID33 with a luminosity of $1
\times 10^{39}$~ergs~s$^{-1}$ at the dynamical center of
NGC~2146 (R.A.=\timeform{06h18m37s.4} $\pm$ \timeform{0s.4},
Dec.=\timeform{78D21'23".2} $\pm$ \timeform{2"}) derived
from a $^{12}$CO rotation curve \citep{Tarchi2000}.
Since the total counts of ID33 is only 113, we cannot make a complete
spectral analysis. However, this source is a hard source (HR $\sim 0.55$)
and the nearby source ID35 is also hard (HR $\sim 0.58$).
As the dense dust lane\citep{Young1988} is across these sources,
it is natural to consider that they are strongly absorbed nuclear
sources. The absorption column density under the assumption of a canonical
AGN power-law spectral model ($\Gamma=2.0$)
is comparable with the value obtained by radio observations \citep{Tarchi2004}.
ID33 may be a low-luminosity AGN (LLAGN). We note that the X-ray observations
with Chandra have first exposed the nuclear region even
under the heavy absorption.

\subsubsection{The $\log{N}-\log{S}$ distribution and Luminosity Function}
\begin{figure}
  \begin{center}
    \FigureFile(60mm,60mm){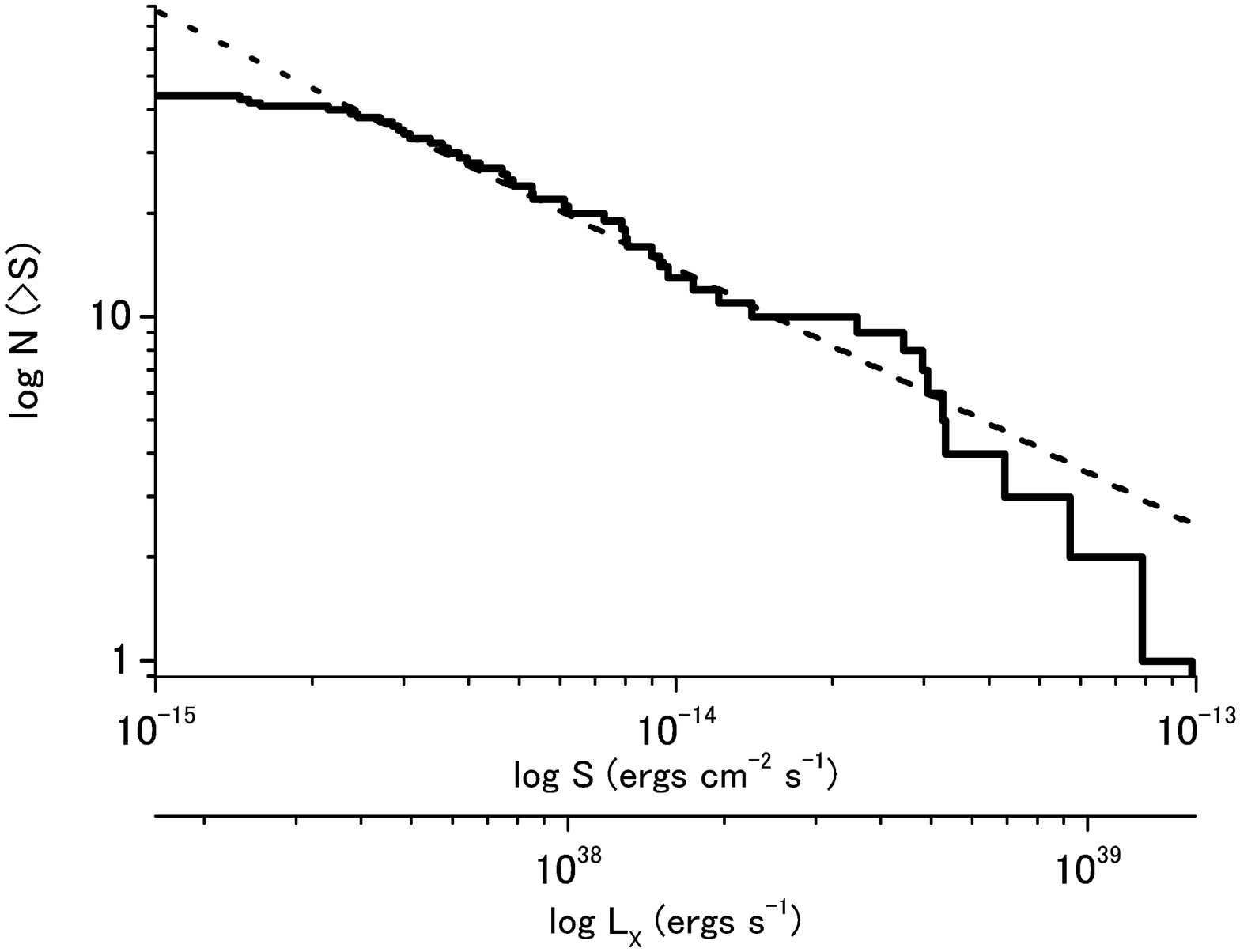}
    %%% \FigureFile(width,height){filename}
  \end{center}
  \caption{
$\log{N}-\log{S}$ distribution from NGC~2146. $S$ is the
observed 2.0-10.0 keV flux. The dashed line is the best-fit
power-law model. The luminosity is calculated by 
$L_{\rm X}=4\pi D^2\cdot S$, where $D$ is the distance to NGC~2146.
The absorption has a negligible effect in the high
energy band.}
\label{FIG:LOGN-LOGS}
\end{figure}

Figure~\ref{FIG:LOGN-LOGS} shows the $\log{N}-\log{S}$
distribution for NGC~2146, where $S$ is the observed 2.0-10.0 keV
flux. The $\log{N}-\log{S}$ has a sharp break at $\sim
2 \times 10^{-15}$~ergs~cm$^{-2}$~s$^{-1}$, that might be
due to an observational detection limit. In order to investigate this
hypothesis, we have estimated an approximate possible value for
such a limit. Within \timeform{0'.5} from the center of NGC~2146,
where the surface brightness of the diffuse emission is
quite large, the typical counts due to the diffuse emission
integrated for 60~ks within the typical PSF size
(\timeform{0".5}) is three counts. Therefore, if a source in
this region has more than 12 counts, which corresponds to
$\sim 1.3 \times 10^{-15}$~ergs~cm$^{-2}$~s$^{-1}$, will be
detected with high significance (more than 5~$\sigma$).
Since the flux is very close to the break point,
we conclude that this break is probably due to the detection limit.

In the total band image, the flux of the faintest source in the
central high diffuse emitting
region is $\sim 2.0\times 10^{-15}$~ergs~cm$^{-2}$~s$^{-1}$.
In the following discussion, we take this value as a conservative
detection limit.

We also derive a luminosity function (LF) for NGC~2146 using the
absorption-corrected 0.5--10.0 keV luminosities. The LF slope
is 0.71. Hence, the LF for NGC~2146 is flatter than that of the
normal spiral galaxies, while it is consistent with those of the other
starburst galaxies \citep{Kilgard2002}. The fact indicates
that the starburst galaxy NGC~2146, similarly to other starburst galaxies,
has a large fraction of high X-ray luminosity
sources.

\subsubsection{Combined point source spectra}
\begin{figure}
  \begin{center}
    \FigureFile(60mm,60mm){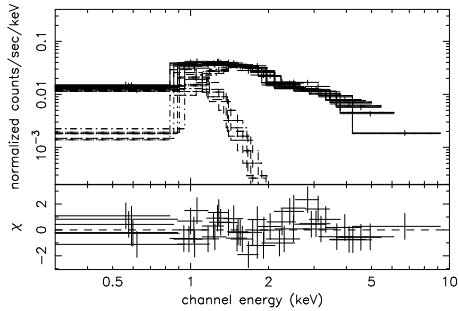}
    %%% \FigureFile(width,height){filename}
  \caption{ Combined spectra of point sources with
fluxes larger than $2\times
10^{-15}$~ergs~cm$^{-2}$~s$^{-1}$ with the two-temperature
bremsstrahlung model.}
\label{FIG:PSRCBB}
  \end{center}
\end{figure}
\begin{table}
\begin{center}
\caption{Results of the spectral fits of the combined point sources}
\label{TBL:PSRCBB}
\begin{tabular}{lc}
\hline\hline
$N_{\rm H}$ ($10^{21} {\rm cm}^{-2}$) & 9.9 (9.6 -- 10.9) \\
$kT_{\rm brems1}$ (keV) & 0.14 (0.12 -- 0.16) \\
$F_{\rm brems1}$\footnotemark[$\dagger$] ($10^{-13}$ ergs cm$^{-2}$ s$^{-1}$) & 0.43 \\
$kT_{\rm brems2}$ (keV) & 6.7 (5.8 -- 9.0) \\
$F_{\rm brems2}$\footnotemark[$\dagger$] ($10^{-13}$ ergs cm$^{-2}$ s$^{-1}$) & 9.0 \\
\hline      
$\chi^2$/dof & 39.4/35 \\
\hline
\multicolumn{2}{@{}l@{}}{\hbox to 0pt{\parbox{90mm}{\footnotesize
       Note---Parentheses indicate the 90\% confidence limit.
       \par\noindent
       \footnotemark[$\dagger$] Observed fluxes in the 0.5--10.0~keV band.
     }\hss}}
\end{tabular}
\end{center}
\end{table}

We combine the spectra of the point sources with
fluxes larger than the detection limit and fitted them by
two-temperature bremsstrahlung model with all parameters
free, since a power-law model, a bremsstrahlung or 
its combination
models cannot describe the spectra. Spectra and the
best-fit model are displayed in Figure~\ref{FIG:PSRCBB}. The
best-fit parameters are listed in Table~\ref{TBL:PSRCBB}.
The total observed flux of point sources in the 2.0--10~keV band is 
$7.8 \times 10^{-13}$~ergs~cm$^{-2}$~s$^{-1}$. Since the total flux
coming from NGC~2146 in the same band is
$9.8 \times 10^{-13}$~ergs~cm$^{-2}$~s$^{-1}$, about 80\% of
the hard emission resides in the point sources, resolved by the high
spatial resolution of Chandra. Table~\ref{TBL:ERP} shows the 2--10~keV
emission ratios of detected point sources by Chandra to the total for
six starburst galaxies. The hard emission of the starburst galaxies can be
divided into two groups; point source dominant and diffuse source dominant.

\begin{table}
\begin{center}
\caption{The 2--10~keV emission ratio of point sources to the total}
\label{TBL:ERP}
\begin{tabular}{lc}
\hline\hline
\multicolumn{2}{c}{point source dominant group}\\
\hline
NGC 2146 & 0.80 \\
M82 & 0.70 \\
the Antenna & 0.81 \\
\hline
\multicolumn{2}{c}{diffuse source dominant group}\\
\hline
NGC 253 & 0.48 \\
NGC 5253 & 0.38 \\
NGC 3256 & 0.20 \\
\hline
\multicolumn{2}{@{}l@{}}{\hbox to 0pt{\parbox{90mm}{\footnotesize
       Reference to \citep{Griffiths2000} for M82,\\
       \citep{Fabbiano2001} for the Antenna,\\
       \citep{Weaver2002} for NGC 253,\\
       \citep{Summers2004} for MGC 5253,\\
       \citep{Lira2002} for NGC 3256.\\
     }\hss}}
\end{tabular}
\end{center}
\end{table}

\subsection{Diffuse Emission}

\subsubsection{Hard Component}

The hard spectral component of the diffuse emission
constitutes 20\% of the hard (2.0--10.0~keV) emission produced in the
whole galaxy, with a luminosity of $\sim 4\times 10^{39}$~ergs~s$^{-1}$.
This luminosity is greater than the hard diffuse emission
of M82 ($\sim 2\times 10^{39}$ ergs s$^{-1}$; \cite{Griffiths2000})
and NGC~253 ($\sim 1\times 10^{39}$~ergs~s$^{-1}$;
\cite{Pietsch2001}. Considered as
a thermal plasma, the temperature is $\sim$ 2.4~keV
(Table~\ref{TBL:VMEKAL}). These features are quite similar
to M82. To determine whether the origin of the hard
component is a hot diffuse gas or a superposition of point
sources unresolved even with Chandra, we estimate the
contributions of the unresolved X-ray point sources.

\begin{figure}
  \begin{center}
    \FigureFile(60mm,60mm){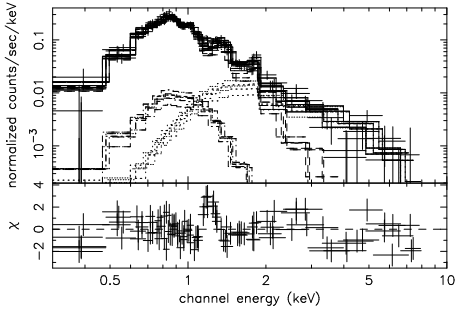}
    %%% \FigureFile(width,height){filename}
  \caption{ Spectra of the diffuse source with the vmekal model plus the unresolved sources model.}\label{FIG:DIF-PSRCBB}
  \end{center}
\end{figure}

At first, we try a spectral approach. We assume that the
shape of the spectrum produced by a number of unresolved sources can be
described through the same model used for the combined spectra
of the resolved sources. Hence, we fit the diffuse emission in the
same way described in \S{4.2.5}. The model implying a collection of 
unresolved sources cannot fit the observed spectra, originating
large residuals below 1~keV. Hence, we add the vmekal model and try
again the fitting procedure and get an acceptable fit.
The best-fit model are shown in
Figure~\ref{FIG:DIF-PSRCBB} and the best-fit parameters are
listed in Table~\ref{TBL:REFIT-VMEKAL}. The observed flux
of the unresolved source component in the 2.0--10.0 keV band
is $3.2 \times 10^{-13}$~ergs~cm$^{-2}$~s$^{-1}$.

\begin{table*}
\begin{center}
\caption{Results of the spectral fits of the diffuse emission with the unresolved sources component}\label{TBL:REFIT-VMEKAL}
\begin{tabular}{lc}
\hline\hline
\multicolumn{2}{c}{thermal plasma (absorbed vmekal)} \\
\hline
$N_{\rm H}$ ($10^{22} {\rm cm}^{-2}$) & 0.15 (0.12 -- 0.20) \\
$kT_{\rm soft}$ (keV) & 0.54 (0.50 -- 0.57) \\
Mg\footnotemark[$*$] & 1.9 (1.6 -- 2.3) \\
Si\footnotemark[$*$] & 2.0 (1.4 -- 2.8) \\
Fe\footnotemark[$*$] & 0.39 (0.34 -- 0.45) \\
$F_{\rm 0.5-2.0keV}$\footnotemark[$\dagger$] ($10^{-13}$ ergs cm$^{-2}$ s$^{-1}$) & 4.9 \\
$L_{\rm 0.5-2.0keV}$\footnotemark[$\dagger$] ($10^{40}$ ergs s$^{-1}$) & 1.3 \\
\hline
\multicolumn{2}{c}{unresolved sources (absorbed two-temperature bremsstrahlung)}\\
\hline
$N_{\rm H}$ ($10^{22} {\rm cm}^{-2}$) & 0.99 (fixed) \\
$kT_{\rm brems1}$ (keV) & 0.14 (fixed) \\
$kT_{\rm brems2}$ (keV) & 6.7 (fixed) \\
$F_{\rm 2.0-10.0keV}$\footnotemark[$\ddagger$] ($10^{-13}$ ergs cm$^{-2}$ s$^{-1}$) & 3.2 \\
$L_{\rm 2.0-10.0keV}$\footnotemark[$\ddagger$] ($10^{40}$ ergs s$^{-1}$) & 0.57 \\
\hline      
$\chi^2$/dof & 138/92 \\
\hline
\multicolumn{2}{@{}l@{}}{\hbox to 0pt{\parbox{185mm}{\footnotesize
       Note---Parentheses indicate the 90\% confidence limit.
	\par\noindent
       \footnotemark[$*$] Abundances are relative to the \cite{Anders1989} solar abundance.
	\par\noindent
       \footnotemark[$\dagger$] Observed flux and absorption-corrected luminosity of the soft component in the 0.5-2.0 keV band.
	\par\noindent
       \footnotemark[$\ddagger$] Observed flux and absorption-corrected luminosity of the hard component in the 2.0-10.0 keV band.
     }\hss}}
\end{tabular}
\end{center}
\end{table*}

Next, we estimate the contributions of the unresolved X-ray
sources by the $\log{N}-\log{S}$ distribution. The best-fit
model function of the $\log{N}-\log{S}$ above $2 \times
10^{-15}$~erg~cm$^{-2}$~s$^{-1}$ can be described as
{\small
\begin{eqnarray}
N(>S) && = \int_{S}^{\infty}n(S)dS\nonumber\\
 && = 49 \left(\frac{S}{2 \times 10^{-15}{\rm~ergs~cm}^{-2} {\rm~s}^{-1}} \right)^{-0.75} {\rm sources},
\end{eqnarray}
}
where $n(S)dS$ is the number of the sources with the flux $S$.
By extrapolating the $\log{N}-\log{S}$ distribution below $2
\times 10^{-15}$~erg~cm$^{-2}$~s$^{-1}$, we can calculate
the total flux $F$ of the sources having the flux between
$S_1$ and $S_2$ by
{\tiny
\begin{eqnarray}
F && =\int_{S_1}^{S_2}S\cdot n(S)dS\nonumber\\
 && = 2.9 \times 10^{-13} \left[\left(\frac{S}{2 \times 10^{-15}{\rm~erg~cm}^{-2} {\rm~s}^{-1} }\right)^{0.25} \right]_{S_1}^{S_2} {\rm~ergs~cm}^{-2} {\rm~s}^{-1}.
\end{eqnarray}
}
According to the equation (2), the flux of the unresolved
sources between 0 and $2\times
10^{-15}$~ergs~cm$^{-2}$~s$^{-1}$ is $F_{\rm unres}= 2.9
\times 10^{-13}$~ergs~cm$^{-2}$~s$^{-1}$, which is very
close to the flux derived by the spectral
approach. Therefore we conclude that the hard emission of
NGC~2146 is probably due to a superposition of unresolved
point sources.

\subsubsection{Soft Component}

As shown in Figure~\ref{FIG:DIF-PSRCBB}, after having fitted the spectra
clear residuals between 1.0 and 1.3~keV are still present.
When we fitted this residual feature using a
Gaussian function, we obtain a value for the feature's central position and
width (FWHM) corresponding to $1.22_{-0.02}^{+0.01}$~keV
and $\le 96$~eV, respectively. Since this feature is 
present at all epochs, we consider it as real.
In this energy range, there are Ni emission
lines. However the features cannot be accounted for by
increasing the Ni abundance because Ni emission lines also exist
at other energies.  We tried to change the gain of ACIS with
-22~eV, but the statistics was not improved
$\chi^2=120/91$(dof). If we replace the soft component with
a two-temperature thermal plasma model, statistics becomes
$\chi^2=116/86$(dof), but this is not an acceptable fit
yet.
This may indicate a neutral Mg emission line (1.25~keV).
This problem will be resolved by the high energy resolution of
Astro-E2 XRS\footnote{See http://www.isas.jaxa.jp/e/enterp/missions/astro-e2/}.
We then estimated the physical parameters
simply by using the parameters reported in
Table~\ref{TBL:REFIT-VMEKAL}.

The plasma temperature is determined to be $kT \sim
0.5$~keV. The plasma has the high abundances for Mg and Si.
The emission integral (EI=$\int n_e^2 dV$) was calculated to
be $8.0 \times 10^{62}$~cm$^{-3}$. We assumed that the total
volume of NGC~2146 to be a sphere with a radius of
\timeform{1'.8}(=6.3~kpc), and we parameterized the
clumpiness of the plasma by a volume(V)-filling factor,
$f$. From these values, we determined the plasma density
[$n_e\sim({\rm EI}/Vf)^{1/2}$], the plasma pressure ($p\sim
2n_ekT$), the plasma mass ($M\sim n_em_pVf$), and the
thermal energy of the plasma ($E\sim 3n_ekTVf$). These physical
parameters are listed in Table~\ref{TBL:SOFTPARAM}. The
origin of the diffuse soft X-ray emission has been discussed
and considered as an outflow gas along the minor axis of
NGC~2146~\citep{Armus1995}.

\begin{table}
\caption{Physical conditions of the soft X-ray-emitting gas}
\label{TBL:SOFTPARAM}
\begin{center}
\begin{tabular}{cc} 
\hline\hline
Parameter & Value \\
\hline
$<n_e>$ & $5.1 \times 10^{-3} f^{-1/2} $cm$^{-3}$\\
$p$ & $4.9 \times 10^{-12} f^{-1/2}$ dyne cm$^{-2}$ \\
$M_{\rm X}$ & $1.3 \times 10^8 f^{1/2} \MO$\\
$E_{\rm X}$ & $2.3 \times 10^{56} f^{1/2}$ ergs \\
\hline
\end{tabular}
\end{center}
\end{table}

\section{Summary}

We observed the starburst galaxy NGC~2146 with the ACIS-S on board Chandra
in six different epochs, with a total exposure time of 60ks. We obtained the
following results:

\begin{enumerate}

\item 
We detected a total of 67 point sources in the ACIS-S field of view and
compiled an X-ray point source catalog for NGC~2146.

\item 
We did not detect any source as luminous as that found in the prototype
starburst galaxy M82 (M82 X-1; luminosity $\sim 10^{41}$~ergs~s$^{-1}$).

\item 
We found no positional coincidence, and hence, any association, between
the detected X-ray point sources and those found in the NIR (2MASS All-Sky PSC)
or radio (MERLIN+VLA observations) band.

\item 
We found a hard X-ray source coincident in position with
the dynamical center of the galaxy. It has a luminosity of
$\sim 1 \times 10^{39}$~ergs~s$^{-1}$ and it may represent
a possible low-luminosity AGN (LLAGN) candidate.

\item 
We derived a $\log{N}-\log{S}$ distribution and a luminosity
function (LF) for NGC~2146. The former shows a break that we
demonstrated to be likely caused by detection limit. The slope of LF
is 0.71, consistent with that of other starburst galaxies. This represents
an indication that NGC~2146, as well as other starbursts, hosts a
larger fraction of luminous sources than normal galaxies.

\item 
We have mapped diffuse emission in both the soft (0.5--2.0 keV)
and hard (2.0--10.0 keV) energy bands. The spectra were
fitted using a two-component model: a soft and a hard one.

\item 
The point sources produce most of the hard emission in NGC~2146 and even
the hard component of the diffuse emission, with a luminosity
of $\sim 4\times 10^{39}$ ergs s$^{-1}$, is probably accounted
for by unresolved point sources.

\item 
The diffuse emission soft component is described by a thermal plasma model
with a temperature of $kT \sim 0.5$~keV with high
abundances for Mg and Si. We also determined the physical
parameters of the soft X-ray-emitting gas.
\end{enumerate}

\section*{Acknowledgement}

We would like to thank the anonymous referee for a considered, helpful and prompt report. This work is supported by the Grant-in-Aid for the 21st Century COE "Center for Diversity and Universality in Physics" from the Ministry of Education, Culture, Sports, Science and Technology (MEXT) of Japan.
This work was supported in part by a grant-in-aid for scientific research 15740119 from the Japan Society for the promotion of Science (JSPS).
Support for this work was provided by the National Aeronautics and
Space Administration through Chandra Award Number GO2-3153X issued
by the Chandra X-ray Observatory Center, which is operated by the
Smithsonian Astrophysical Observatory for and on behalf of the
National Aeronautics and Space Administration uder contract
NAS8-39073.

%%%%%%%%%%%%%%%%%%%%%%%%%%%%%%%%%%%%%%%

%%%
% See the manual for the detail.
%%%

\begin{landscape}
\begin{table*}
%\begin{center}
\vspace{-1.8cm}
\caption{Main Chandra Point Source Catalog}\label{TBL:CATALOG}
\hspace{-7.0cm}
\tiny
\begin{tabular}{ccccccccccccc}
\hline\hline
 & R.A. & Dec. & Net &  & Time & Hardness &  &  &  &  &  &  \\
ID & (J2000) & (J2000) & Counts & Significance & Variability & Ratio & $N_{H}$ & $\Gamma$ & $F_{\rm X,2-10}$ & $F_{\rm X,0.5-10}$ & $\log{L_{\rm X,2-10}}$ & $\log{L_{\rm X,0.5-10}}$ \\
 &  &  & (1) & (2) & (3) & (4) & (5) & (6) & (7) & (8) & (9) & (10) \\\hline
1 & \timeform{06h17m21s.69} & \timeform{+78D22'21".3} & 339.6 $\pm$ 18.8 & 67.6 & 0.95 & -0.43 (-0.49 -- -0.38) & 0.10 (... -- 0.36) & 1.62 (1.17 -- 2.76) & 3.30 & 4.56 & 38.73 & 38.90 \\
2 & \timeform{06h17m21s.73} & \timeform{+78D17'43".7} & 8.9 $\pm$ 3.3 & 3.3 & 0.05 & -0.66 (... -- -0.42) & \footnotemark[$b$] & \footnotemark[$a$] & 0.08 & 0.10 & 37.11 & 37.38 \\
3 & \timeform{06h17m23s.64} & \timeform{+78D20'36".3} & 36.4 $\pm$ 6.3 & 11.8 & 0.42 & -0.83 (-0.94 -- -0.72) & \footnotemark[$b$] & \footnotemark[$a$] & 0.31 & 0.42 & 37.71 & 37.98 \\
4 & \timeform{06h17m23s.88} & \timeform{+78D20'33".9} & 61.2 $\pm$ 8.1 & 18.3 & 0.74 & -0.75 (-0.85 -- -0.65) & \footnotemark[$b$] & \footnotemark[$a$] & 0.47 & 0.65 & 37.90 & 38.17 \\
5 & \timeform{06h17m40s.68} & \timeform{+78D22'16".0} & 40.3 $\pm$ 6.7 & 12.0 & 0.28 & -0.62 (-0.77 -- -0.50) & \footnotemark[$b$] & \footnotemark[$a$] & 0.38 & 0.52 & 37.81 & 38.08 \\
6 & \timeform{06h17m48s.59} & \timeform{+78D24'14".7} & 65.2 $\pm$ 8.3 & 20.9 & 0.75 & -0.48 (-0.61 -- -0.37) & \footnotemark[$b$] & \footnotemark[$a$] & 0.77 & 1.05 & 38.11 & 38.38 \\
7 & \timeform{06h17m51s.58} & \timeform{+78D25'35".4} & 24.3 $\pm$ 5.3 & 7.8 & 0.02 &  0.00 (-0.26 -- 0.19) & \footnotemark[$b$] & \footnotemark[$a$] & 0.29 & 0.39 & 37.68 & 37.95 \\
8 & \timeform{06h17m56s.48} & \timeform{+78D24'11".7} & 6.5 $\pm$ 2.6  & 3.1 & 0.00 & -0.48 (... -- -0.16) & \footnotemark[$b$] & \footnotemark[$a$] & 0.09 & 0.13 & 37.19 & 37.46 \\
9 & \timeform{06h18m04s.24} & \timeform{+78D23'54".3} & 8.5 $\pm$ 3.0 & 4.1 & 0.03 & -0.18 (-0.89 -- 0.18) & \footnotemark[$b$] & \footnotemark[$a$] & 0.09 & 0.12 & 37.18 & 37.45 \\
10 & \timeform{06h18m04s.60} & \timeform{+78D19'16".6} & 18.6 $\pm$ 4.8 & 5.7 & 0.08 & -0.27 (-0.54 -- -0.08) & \footnotemark[$b$] & \footnotemark[$a$] & 0.20 & 0.27 & 37.52 & 37.79 \\
11 & \timeform{06h18m05s.40} & \timeform{+78D23'10".7} & 58.3 $\pm$ 7.8 & 20.4 & 0.04 & 0.17 (0.02 -- 0.30) & \footnotemark[$b$] & \footnotemark[$a$] & 0.99 & 1.35 & 38.22 & 38.49 \\
12 & \timeform{06h18m06s.13} & \timeform{+78D25'38".8} & 11.2 $\pm$ 3.6 & 4.3 & 0.02 & 0.36 (0.00 -- 0.60) & \footnotemark[$b$] & \footnotemark[$a$] & 0.18 & 0.25 & 37.52 & 37.79 \\
13 & \timeform{06h18m09s.46} & \timeform{+78D21'25".3} & 15.1 $\pm$ 4.5 & 4.5 & 0.15 & -0.40 (-0.71 -- -0.19) & \footnotemark[$b$] & \footnotemark[$a$] & 0.19 & 0.26 & 37.50 & 37.77 \\
14 & \timeform{06h18m13s.01} & \timeform{+78D21'50".7} & 40.4 $\pm$ 6.7 & 12.2 & 0.14 & -0.74 (-0.86 --  -0.65) & \footnotemark[$b$] & \footnotemark[$a$] & 0.46 & 0.63 & 37.89 & 38.16 \\
15 & \timeform{06h18m13s.06} & \timeform{+78D22'43".3} & 109.1 $\pm$ 10.8 & 18.9 & 0.80 & -0.50 (-0.60 -- -0.42) & \footnotemark[$b$] & \footnotemark[$a$] & 1.08 & 1.47 & 38.26 & 38.53 \\
16 & \timeform{06h18m16s.30} & \timeform{+78D21'34".9} & 10.1 $\pm$ 3.7 & 3.2 & 0.12 & 0.03 (-0.36 -- 0.28) & \footnotemark[$b$] & \footnotemark[$a$] & 0.12 & 0.16 & 37.30 & 37.57 \\
17 & \timeform{06h18m17s.63} & \timeform{+78D20'17".7} & 15.3 $\pm$ 4.2 & 5.4 & 0.42 & -0.58 (-0.88 -- -0.37) & \footnotemark[$b$] & \footnotemark[$a$] & 0.15 & 0.21 & 37.40 & 37.67 \\
18 & \timeform{06h18m17s.84} & \timeform{+78D22'26".9} & 76.4 $\pm$ 9.1 & 21.7 & 0.93 & -0.36 (-0.48 -- -0.26) & \footnotemark[$b$] & \footnotemark[$a$] & 0.81 & 1.10 & 38.13 & 38.40 \\
19 & \timeform{06h18m21s.83} & \timeform{+78D24'56".6} & 7.2 $\pm$ 2.8  & 3.2  & 0.04 & -0.76 (... -- -0.51) & \footnotemark[$b$] & \footnotemark[$a$] & 0.15 & 0.20 & 37.39 & 37.66 \\
20 & \timeform{06h18m23s.81} & \timeform{+78D22'25".0} & 17.2 $\pm$ 4.5 & 5.9 & 0.04 & -0.16 (-0.49 -- 0.06) & \footnotemark[$b$] & \footnotemark[$a$] & 0.16 & 0.22 & 37.43 & 37.70 \\
21 & \timeform{06h18m30s.57} & \timeform{+78D21'31".8} & 39.2 $\pm$ 7.5 & 7.6 & 0.37 & -0.28 (-0.47 -- -0.14) & \footnotemark[$b$] & \footnotemark[$a$] & 0.42 & 0.57 & 37.85 & 38.12 \\
22 & \timeform{06h18m30s.86} & \timeform{+78D21'22".8} & 56.6 $\pm$ 9.0  & 9.4 & 0.44 & -0.06 (-0.22 -- 0.07) & \footnotemark[$b$] & \footnotemark[$a$] & 0.73 & 0.99 & 38.09 & 38.36 \\
23 & \timeform{06h18m30s.86} & \timeform{+78D22'43".2} & 56.6 $\pm$ 7.9 & 15.4 & 0.81 & -0.56 (-0.69 -- -0.45) & \footnotemark[$b$] & \footnotemark[$a$] & 0.53 & 0.72 & 37.95 & 38.22 \\
24 & \timeform{06h18m31s.78} & \timeform{+78D21'41".4} & 67.9 $\pm$ 9.1 & 13.6 & 0.21 & 0.58 (0.48 -- 0.67) & \footnotemark[$b$] & \footnotemark[$a$] & 0.62 & 0.85 & 38.02 & 38.29 \\
25 & \timeform{06h18m32s.27} & \timeform{+78D21'31".0} & 28.2 $\pm$ 6.6 & 5.6 & 0.03 & -0.68 (-0.87 -- -0.52) & \footnotemark[$b$] & \footnotemark[$a$] & 0.36 & 0.49 & 37.78 & 38.05 \\
26 & \timeform{06h18m32s.77} & \timeform{+78D21'22".6} & 32.4 $\pm$ 7.2 & 5.9 & 0.33 & -0.49 (-0.68 -- -0.34) & \footnotemark[$b$] & \footnotemark[$a$] & 0.37 & 0.50 & 37.79 & 38.06 \\
27 & \timeform{06h18m33s.61} & \timeform{+78D20'43".1} & 50.5 $\pm$ 7.5 & 14.4 & 0.60 & -0.16 (-0.32 -- -0.03) & \footnotemark[$b$] & \footnotemark[$a$] & 0.49 & 0.66 & 37.91 & 38.18 \\
28 & \timeform{06h18m34s.62} & \timeform{+78D21'25".4} & 25.5 $\pm$ 6.3 & 5.2 & 0.28 & 0.53 (0.36 -- 0.66) & \footnotemark[$b$] & \footnotemark[$a$] & 0.29 & 0.39 & 37.68 & 37.95 \\
29 & \timeform{06h18m34s.65} & \timeform{+78D21'33".7} & 124.0 $\pm$ 12.3 & 20.0 & 1.00 & 0.57 (0.49 -- 0.64) & 2.30 (1.67 -- 3.06) & \footnotemark[$a$] & 3.26 & 3.49 & 38.81 & 39.08 \\
30 & \timeform{06h18m36s.71} & \timeform{+78D21'34".0} & 32.5 $\pm$ 7.1 & 6.2 & 0.01 & -1.27\footnotemark[$c$] & \footnotemark[$b$] & \footnotemark[$a$] & 0.53 & 0.72 & 37.95 & 38.22 \\
31 & \timeform{06h18m36s.71} & \timeform{+78D20'48".7} & 23.9 $\pm$ 5.6 & 6.3 & 0.27 & -0.44 (-0.68 -- -0.26) & \footnotemark[$b$] & \footnotemark[$a$] & 0.25 & 0.33 & 37.61 & 37.88 \\
32 & \timeform{06h18m36s.90} & \timeform{+78D21'27".4} & 349.9 $\pm$ 10.3 & 38.9 & 0.99 & 0.44 (0.39 --  0.48) & 2.95 (1.22 -- 5.73) & 2.23 (1.21 -- 3.53) & 9.82 & 10.42 & 39.32 & 39.66 \\
33 & \timeform{06h18m37s.30} & \timeform{+78D21'23".0} & 112.9 $\pm$ 12.4  & 15.3  & 0.33 & 0.55 (0.47 -- 0.62) & 2.04 (1.33 -- 3.01) & \footnotemark[$a$] & 3.05 & 3.31 & 38.77 & 39.04 \\
34 & \timeform{06h18m37s.32} & \timeform{+78D21'14".4} & 16.8 $\pm$ 5.2 & 3.9 & 0.03 & -2.60\footnotemark[$c$] & \footnotemark[$b$] & \footnotemark[$a$] & 0.21 & 0.28 & 37.54 & 37.81 \\
35 & \timeform{06h18m37s.47} & \timeform{+78D21'20".4} & 88.0 $\pm$ 11.6 & 11.1 & 0.68 & 0.58 (0.50 -- 0.66) & 2.35 (1.64 -- 3.39) & \footnotemark[$a$] & 2.98 & 3.18 & 38.77 & 39.04 \\
36 & \timeform{06h18m37s.57} & \timeform{+78D20'21".4} & 31.1 $\pm$ 5.9 & 9.8 & 0.17 & -0.36 (-0.56 -- -0.20) & \footnotemark[$b$] & \footnotemark[$a$] & 0.29 & 0.40 & 37.69 & 37.93 \\
37 & \timeform{06h18m37s.72} & \timeform{+78D21'24".9} & 229.3 $\pm$ 16.8 & 27.4 & 0.22 & 0.18 (0.11 --  0.24) & 1.12 (... -- ...) & 1.94 (... -- ...) & 4.29 & 4.96 & 38.89 & 39.14 \\
38 & \timeform{06h18m37s.90} & \timeform{+78D21'11".0} & 32.2 $\pm$ 7.1 & 6.1 & 0.02 & -0.59 (-0.78 -- -0.45) & \footnotemark[$b$] & \footnotemark[$a$] & 0.40 & 0.54 & 37.82 & 38.09 \\
39 & \timeform{06h18m38s.62} & \timeform{+78D21'27".4} & 117.4 $\pm$ 12.5 & 16.1 & 0.40 & -0.21 (-0.31 -- -0.12) & 0.25 (0.06 -- 0.68) & \footnotemark[$a$] & 0.97 & 1.42 & 38.20 & 38.47 \\
40 & \timeform{06h18m38s.77} & \timeform{+78D21'24".3} & 214.3 $\pm$ 17.1 & 21.5 & 0.38 & -0.21 (-0.29 -- -0.15) & 0.31 (... -- 0.77) & 1.86 (1.01 -- 2.94) & 2.23 & 3.01 & 38.57 & 38.80 \\
41 & \timeform{06h18m39s.31} & \timeform{+78D21'21".5} & 46.7 $\pm$ 7.5 & 10.8 & 0.66 & 0.62 (0.51 -- 0.72) & \footnotemark[$b$] & \footnotemark[$a$] & 0.90 & 1.23 & 38.18 & 38.45 \\
42 & \timeform{06h18m39s.61} & \timeform{+78D21'14".5} & 568.0 $\pm$ 24.9 & 68.7 & 0.96 & -0.14 (-0.19 -- -0.10) & 1.08 (0.71 -- 1.52) & 2.33 (2.06 -- 2.89) & 7.88 & 9.83 & 39.16 & 39.54 \\
43 & \timeform{06h18m40s.04} & \timeform{+78D21'16".8} & 77.4 $\pm$ 10.2 & 12.2 & 0.27 & 0.63 (0.53 -- 0.72) & \footnotemark[$b$] & \footnotemark[$a$] & 0.61 & 0.83 & 38.01 & 38.28 \\
44 & \timeform{06h18m40s.87} & \timeform{+78D21'24".8} & 273.3 $\pm$ 18.7 & 28.2 & 0.43 & -0.71 (-0.77 -- -0.66) & 0.26 (0.14 -- 0.44) & \footnotemark[$a$] & 1.40 & 2.03 & 38.36 & 38.63 \\
45 & \timeform{06h18m41s.10} & \timeform{+78D23'47".0} & 8.3 $\pm$ 3.0 & 3.8 & 0.01 & 0.10 (-0.43 -- 0.40) & \footnotemark[$b$] & \footnotemark[$a$] & 0.14 & 0.20 & 37.39 & 37.66 \\
46 & \timeform{06h18m41s.49} & \timeform{+78D21'14".0} & 132.6 $\pm$ 13.4 & 16.7 & 0.37 & -0.39 (-0.47 -- -0.31) & 0.26 (0.07 -- 0.47) & \footnotemark[$a$] & 1.21 & 1.77 & 38.30 & 38.57 \\
47 & \timeform{06h18m42s.40} & \timeform{+78D21'13".9} & 149.2 $\pm$ 13.8 & 20.0 & 0.65 & 0.14 (0.05 --  0.22) & 1.19 (0.80 -- 1.69) & \footnotemark[$a$] & 2.74 & 3.17 & 38.69 & 38.96 \\
48 & \timeform{06h18m43s.41} & \timeform{+78D21'21".1} & 66.3 $\pm$ 10.0 & 9.8 & 0.02 & -0.29 (-0.43 --  -0.17) & \footnotemark[$b$] & \footnotemark[$a$] & 0.80 & 1.09 & 38.13 & 38.40 \\
49 & \timeform{06h18m45s.64} & \timeform{+78D20'43".1} & 22.4 $\pm$ 5.2 & 6.8 & 0.38 & -0.09 (-0.34 -- 0.09) & \footnotemark[$b$] & \footnotemark[$a$] & 0.27 & 0.37 & 37.66 & 37.93 \\
50 & \timeform{06h18m46s.58} & \timeform{+78D21'00".8} & 112.1 $\pm$ 11.3 & 22.6 & 0.97 & -0.50 (-0.59 -- -0.42) & 0.31 (0.17 -- 0.49) & \footnotemark[$a$] & 0.93 & 1.32 & 38.19 & 38.46 \\
51 & \timeform{06h18m48s.65} & \timeform{+78D25'06".1} & 257.6 $\pm$ 16.3 & 66.8 & 0.93 & -0.47 (-0.53 -- -0.41) & 0.34 (0.11 -- 0.52) & 2.63 (1.97 -- 3.71) & 1.33 & 2.48 & 38.35 & 38.85 \\
52 & \timeform{06h18m49s.92} & \timeform{+78D20'45".4} & 15.3 $\pm$ 4.4 & 4.9 & 0.00 & -0.14 (-0.43 -- 0.07) & \footnotemark[$b$] & \footnotemark[$a$] & 0.18 & 0.25 & 37.49 & 37.76 \\
53 & \timeform{06h18m51s.01} & \timeform{+78D20'58".6} & 26.8 $\pm$ 5.7 & 7.8 & 0.13 & -0.93 (... -- -0.87) & \footnotemark[$b$] & \footnotemark[$a$] & 0.21 & 0.29 & 37.56 & 37.83 \\
54 & \timeform{06h19m00s.74} & \timeform{+78D21'09".0} & 22.1 $\pm$ 5.2 & 6.5 & 0.12 & 0.66 (0.47 -- 0.81) & \footnotemark[$b$] & \footnotemark[$a$] & 0.24 & 0.32 & 37.60 & 37.87 \\
55 & \timeform{06h19m05s.48} & \timeform{+78D19'51".7} & 125.0 $\pm$ 11.4 & 34.6 & 0.47 & -0.60 (-0.68 -- -0.52) & 0.15 (0.05 -- 0.27) & \footnotemark[$a$] & 0.79 & 1.24 & 38.11 & 38.38 \\
56 & \timeform{06h19m07s.32} & \timeform{+78D21'09".0} & 8.3 $\pm$ 3.2 & 3.2 & 0.07 & -0.62 (-0.95 -- -0.39) & \footnotemark[$b$] & \footnotemark[$a$] & 0.05 & 0.07 & 36.95 & 37.22 \\
57 & \timeform{06h19m08s.54} & \timeform{+78D20'25".1} & 15.9 $\pm$ 4.4 & 5.4 & 0.13 & -0.28 (-0.60 -- -0.06) & \footnotemark[$b$] & \footnotemark[$a$] & 0.14 & 0.20 & 37.39 & 37.66 \\
58 & \timeform{06h19m09s.74} & \timeform{+78D19'49".9} & 31.6 $\pm$ 6.0 & 9.7 & 0.01 & -0.06 (-0.27 -- 0.10) & \footnotemark[$b$] & \footnotemark[$a$] & 0.34 & 0.46 & 37.75 & 38.74 \\
59 & \timeform{06h19m10s.00} & \timeform{+78D19'12".5} & 38.2 $\pm$ 6.5 & 12.1 & 0.01 & 0.95 (0.90 -- ...) & \footnotemark[$b$] & \footnotemark[$a$] & 0.15 & 0.21 & 37.40 & 37.47 \\
60 & \timeform{06h19m13s.98} & \timeform{+78D21'45".0} & 391.4 $\pm$ 20.0 & 92.8 & 0.48 & 0.10 (0.05 --  0.15) & 2.40 (1.17 -- 6.54) & 2.53 (1.19 -- 6.06) & 5.73 & 6.84 & 39.11 & 39.26 \\
61 & \timeform{06h19m16s.72} & \timeform{+78D23'08".7} & 12.0 $\pm$ 3.7 & 4.5 & 0.04 & -0.36 (-0.74 -- -0.12) & \footnotemark[$b$] & \footnotemark[$a$] & 0.11 & 0.16 & 37.28 & 37.55 \\
62 & \timeform{06h19m24s.04} & \timeform{+78D17'41".4} & 53.4 $\pm$ 8.2 & 11.2 & 0.79 & -0.07 (-0.22 -- 0.06) & \footnotemark[$b$] & \footnotemark[$a$] & 0.59 & 0.81 & 38.00 & 38.27 \\
63 & \timeform{06h19m28s.00} & \timeform{+78D22'54".1} & 7.0 $\pm$ 2.8 & 3.0 & 0.00 & 1.00 & \footnotemark[$b$] & \footnotemark[$a$] & 0.01 & 0.02 & 36.29 & 36.56 \\
64 & \timeform{06h19m28s.52} & \timeform{+78D21'58".9} & 28.6 $\pm$ 5.7 & 9.4 & 0.10 & -0.29 (-0.51 -- -0.12) & \footnotemark[$b$] & \footnotemark[$a$] & 0.30 & 0.41 & 37.71 & 37.98 \\
65 & \timeform{06h19m30s.46} & \timeform{+78D24'24".8} & 9.9 $\pm$ 3.3 & 4.2 & 0.00 & -1.00 & \footnotemark[$b$] & \footnotemark[$a$] & 0.06 & 0.08 & 37.01 & 37.28 \\
66 & \timeform{06h19m31s.23} & \timeform{+78D23'27".0} & 109.7 $\pm$ 10.8 & 30.1 & 0.42 & -0.66 (-0.74 -- -0.59) & \footnotemark[$b$] & \footnotemark[$a$] & 1.13 & 1.55 & 38.28 & 38.55 \\
67 & \timeform{06h19m34s.92} & \timeform{+78D23'06".3} & 22.1 $\pm$ 5.0 & 7.6 & 0.06 & 0.03 (-0.26 -- 0.24) & \footnotemark[$b$] & \footnotemark[$a$] & 0.23 & 0.31 & 37.59 & 37.86 \\
\hline
\multicolumn{13}{l}{
	\hbox to 0pt{\parbox{234mm}{\tiny
	Notes---Parenthese indicate the 90\% confidence limit. No value in a parenthise means that error is not constrained.
	Col.(1): Photon counts and errors with background substracted.
	Col.(2): Source significance ($\sigma$).
	Col.(3): Time variability given with the $\chi^2$ test.
	Col.(4): Hardness Ratio (HR) derived by the equation $(h-s)/(h+s)$, $h$ is hard band counts (2.0--10.0 keV) and $s$ is soft band counts (0.5--2.0 keV).
	Col.(5): Absorption column density ($10^{22} \rm cm^{-2}$).
	Col.(6): Photon index.
	Col.(7) \& (8): Observational flux in the 2.0--10.0 keV and the 0.5--10.0 keV band ($10^{-14}$ ergs cm$^{-2}$ s$^{-1}$).
	Col.(9) \& (10): Absorption-corrected luminosity in the 2.0--10.0 keV and the 0.5--10.0 keV band. (ergs s$^{-1}$)
    \par\noindent
	\footnotemark[$a$] Photon index fixed at 2.0 due to low number of counts ($<180$ counts).
	\par\noindent
	\footnotemark[$b$] Absorption column density fixed at 0.4 ($10^{22} \rm cm^{-2}$), the best-fit value of the diffuse emission ($<120$ counts).
	\par\noindent
	\footnotemark[$c$] Since a background-substracted hard band counts drop into minus, HR shows less than -1.
 }}}
\end{tabular}
%\end{center}
\end{table*}
\end{landscape}
\end{document}